\documentclass[twocolumn]{aastex701}
\usepackage{graphicx}
\usepackage{color}
\usepackage{amsmath}
\usepackage{mathtools}
\usepackage{tcolorbox}
\usepackage{amssymb}

\usepackage{gensymb}
\usepackage{tabularx} 
\RequirePackage{snapshot}
\usepackage{listings}

\shorttitle{Binary Microlensing in BAGLE}
\shortauthors{Bhadra et al.}

\definecolor{codegreen}{rgb}{0,0.6,0}
\definecolor{codegray}{rgb}{0.3,0.3,0.3}
\definecolor{codepurple}{rgb}{0.58,0,0.82}
\definecolor{backcolour}{rgb}{0.9,0.9,0.9}

\lstdefinestyle{mystyle}{
    backgroundcolor=\color{backcolour},   
    commentstyle=\color{codegreen},
    keywordstyle=\color{magenta},
    numberstyle=\tiny\color{codegray},
    stringstyle=\color{codepurple},
    basicstyle=\ttfamily\footnotesize,
    breakatwhitespace=false,         
    breaklines=true,                 
    captionpos=b,                    
    keepspaces=true,                 
    numbers=left,                    
    numbersep=5pt,                  
    showspaces=false,                
    showstringspaces=false,
    showtabs=false,                  
    tabsize=2
}

\newcommand{\geotr}{\earth_r}

\newcommand{\vect}[1]{\boldsymbol{#1}}
\newcommand{\murel}{\mu_{\mathrm{rel},\sun}}

\newcommand{\accSsec}{\bold{a}_{\boldsymbol{S\mathrm{rel}},\sun}}
\newcommand{\accLsec}{\bold{a}_{\boldsymbol{L\mathrm{rel}},\sun}}

\newcommand{\murelhat}{\boldsymbol{\hat{\mu}}_{\boldsymbol{\mathrm{rel}},\sun}}

\newcommand{\tnot}{t_{0,\sun}}

\newcommand{\uvecgeotr}{\vect{u}_{\geotr}}

\newcommand{\tE}{t_{E,\sun}}

\newcommand{\tEgeotr}{t_{E,\geotr}}
\newcommand{\thetaE}{\theta_E}
\newcommand{\uo}{u_{0,\sun}}
\newcommand{\uocom}{u_{com, 0,\sun}}

\newcommand{\uogeotr}{u_{0,\geotr}}

\newcommand{\piEvecgeotr}{\vect{\pi}_{E,\geotr}}
\newcommand{\piE}{\pi_E}

\newcommand{\tnotgeotr}{t_{0,\geotr}}
\newcommand{\tpnotgeotr}{t_{prim, 0,\geotr}}
\newcommand{\tsnotgeotr}{t_{sec, 0,\geotr}}

\newcommand{\x}{E}
\newcommand{\y}{N}

\newcommand{\xmark}{}

\newcommand{\musvec}{\vect{\mu}_{\boldsymbol{S},\sun}}
\newcommand{\mussysvec}{\vect{\mu}_{\boldsymbol{S, com},\sun}}
\newcommand{\mulsysvec}{\vect{\mu}_{\boldsymbol{L, com},\sun}}

\newcommand{\mussvec}{\vect{\mu}_{\boldsymbol{S_s},\sun}}
\newcommand{\deltamussvec}{\vect{\Delta\mu}_{\boldsymbol{S_s},\sun}}

\newcommand{\deltamulsvec}{\vect{\Delta\mu}_{\boldsymbol{L_s},\sun}}

\newcommand{\mulvec}{\vect{\mu}_{\boldsymbol{L},\sun}}

\newcommand{\mulsvec}{\vect{\mu}_{\boldsymbol{L_s} ,\sun}}

\newcommand{\Xsvec}{\vect{X}_{\boldsymbol{S},\sun}}
\newcommand{\Xlvec}{\vect{X}_{\boldsymbol{L},\sun}}

\newcommand{\bsff}{b_{sff}}

\newcommand{\Xspovec}{\vect{X}_{\boldsymbol{S_p,0},\sun}}

\newcommand{\Xlpovec}{\vect{X}_{\boldsymbol{L_p,0},\sun}}

\newcommand{\Xscomvec}{\vect{X}_{\boldsymbol{S_{com},0},\sun}}
\newcommand{\Xlcomvec}{\vect{X}_{\boldsymbol{L_{com},0},\sun}}

\newcommand{\Xssovec}{\vect{X}_{\boldsymbol{S_s,0},\sun}}
\newcommand{\Xlsovec}{\vect{X}_{\boldsymbol{L_s,0},\sun}}

\newcommand{\Xspvec}{\vect{X}_{\boldsymbol{S_p},\sun}}

\newcommand{\Xlpvec}{\vect{X}_{\boldsymbol{L_p},\sun}}
\newcommand{\Xlsvec}{\vect{X}_{\boldsymbol{L_s},\sun}}

\newcommand{\Xcomp}{\vect{X(t)}_{\boldsymbol{p},\boldsymbol{com},\sun}}
\newcommand{\Xcoms}{\vect{X(t)}_{\boldsymbol{s},\boldsymbol{com},\sun}}

\newcommand{\Xcomep}{{X(t)}_{{p, {com}, E},\sun}}
\newcommand{\Xcomnp}{{X(t)}_{{p, {com}, N},\sun}}

\newcommand{\Xcomes}{{X(t)}_{{s, {com}, E},\sun}}
\newcommand{\Xcomns}{{X(t)}_{{s, {com}, N},\sun}}

\newcommand{\Xssvec}{\vect{X}_{\boldsymbol{S_s},\sun}}

\newcommand{\upvec}{\vect{u}_{p,\sun}}
\newcommand{\usvec}{\vect{u}_{s,\sun}}

\newcommand{\upveco}{\vect{u}_{p,\boldsymbol{0},\sun}}
\newcommand{\usveco}{\vect{u}_{s,\boldsymbol{0},\sun}}

\newcommand{\upo}{u_{p,0,\sun}}
\newcommand{\uso}{u_{s,0,\sun}}
\newcommand{\upogeotr}{u_{p,0,\geotr}}
\newcommand{\usogeotr}{u_{s,0,\geotr}}

\newcommand{\tpnot}{t_{prim,0,\sun}}
\newcommand{\tcomnot}{t_{com,0,\sun}}
\newcommand{\tsnot}{t_{sec,0,\sun}}

\newcommand{\w}{\omega_{pri}}
\newcommand{\wsec}{\omega_{sec}}
\newcommand{\bigomega}{\Omega_{sec}}

\newcommand{\inclination}{\textit{i}}
\newcommand{\eccentricity}{\textit{e}}
\newcommand{\period}{\textit{P}}

\newcommand{\al}{\aleph_{pri}}
\newcommand{\ala}{\aleph_{sec}}
\newcommand{\E}{\textit{E(t)}}
\newcommand{\M}{\textit{M(t)}}

\newcommand{\etanom}{\eta}

\newcommand{\Apri}{A_{pri}}
\newcommand{\Asec}{A_{sec}}
\newcommand{\Bpri}{B_{pri}}
\newcommand{\Bsec}{B_{sec}}

\newcommand{\Cpri}{C_{pri}}
\newcommand{\Csec}{C_{sec}}
\newcommand{\Fpri}{F_{pri}}
\newcommand{\Fsec}{F_{sec}}
\newcommand{\Gpri}{G_{pri}}
\newcommand{\Gsec}{G_{sec}}
\newcommand{\Hpri}{H_{pri}}
\newcommand{\Hsec}{H_{sec}}

\newcommand{\bagle}{BAGLE}

\newcommand{\berkeley}{\affiliation{Department of Astronomy, University of California, Berkeley, CA 94720, USA}}

\begin{document}

\title{Modeling Binary Lenses and Sources with the BAGLE Python Package}
\author[0009-0002-6097-9030]{T. Dex Bhadra}
\affiliation{Department of Astronomy, University of Maryland, College Park, MD 20742, USA}
\affiliation{Code 667, NASA Goddard Space Flight Center, Greenbelt, MD 20771, USA}
\email[show]{tmbhadra@umd.edu}

\author[0000-0001-9611-0009]{J. R. Lu}
\berkeley
\email{}

\author[0000-0002-0287-3783]{Natasha S. Abrams}
\berkeley
\email{}

\author[0000-0002-1395-5426]{Andrew Scharf}
\affiliation{Department of Mathematics, University of California, Berkeley, CA 94720, USA}
\email{scharfa@berkeley.edu}

\author[0000-0003-0652-1862]{Edward Broadberry} 
\affiliation{Department of Astronomy, University of Maryland, College Park, MD 20742, USA}
\email{edbroad@umd.edu}

\author[0000-0002-6406-1924]{Casey Lam}
\berkeley
\affiliation{Observatories of the Carnegie Institution for Science, Pasadena, CA 91101, USA}
\email{}

\author[0000-0003-4591-3201]{Macy J. Huston}
\berkeley
\email{mhuston@berkeley.edu}

\date{\today}

\begin{abstract}
Gravitational microlensing is a powerful tool that can be used to find and measure the mass of isolated and dark compact objects. In many microlensing events, the lens, the source, or both may be a binary system. In this work, we introduce binary source and lens models into the gravitational lensing formalism encoded in the Bayesian Analysis of Gravitational Lensing Events (BAGLE) Python software package. These new binary models in BAGLE account for Keplerian orbits. We also add binary models with fewer parameters that describe the binary orbital motion as acceleration, linear, or stationary motion of the secondary companion; these are useful when the orbit has a very low eccentricity or the orbital period is much longer than the microlensing timescale. The model parameterizations based on these binary lensing equations enable joint-fitting of photometric and astrometric datasets. These binary models will be used to fit microlensing event data from the Vera C. Rubin Observatory, the Nancy Grace Roman Telescope, and other surveys. 

\end{abstract}

\section{Introduction
\label{sec:introduction}}






Gravitational microlensing occurs in the Milky Way when a foreground object with mass (e.g., a star, black hole, or planet) passes in front of a background source star and the mass of the foreground lens temporarily magnifies and perturbs the observed position of the background source. 
Microlensing is detectable even when the foreground lens is dark or too faint to observe, making it a powerful tool for probing cool, distant, and/or compact objects. It is one of the only methods for measuring the mass of isolated and dark black holes \citep{Lam:2022, Sahu:2022, Mroz:2022, Lam:2023-OB110462, Sahu2025}, free-floating or widely separated low-mass exoplanets \citep{Gaudi2012}, and white dwarfs \citep{Sahu2017, McGill2023}. Additionally, over 200 exoplanets orbiting their host stars have been detected with microlensing \citep{Mroz_2024}.

Microlensing is sensitive to both close and wide separation binaries, unlike radial velocity and transit probes, which are more sensitive to closely separated binaries or planet+star systems.
Recent simulations of Milky Way microlensing surveys show that 55\% of observed microlensing events involve a binary star system \citep{Natasha_2025} most of which can masquerade as single source-single lens events. This is significantly more than the $\sim 5-11\%$ of microlensing events with obvious binaries which are commonly reported \citep[e.g.][]{Alcock:2000-Binaries, Jaroszynski:2002, Mroz:2019}, so robust binary microlensing models are even more important than previously thought. 
Some microlensing events have timescales greater than the orbital period of the involved binary system. In these cases, photometric lightcurves and astrometric trajectories can only be properly modeled by accounting for the system's orbital dynamics \citep{Dominik:1998a, Skowron_2011, Penny_2011, Bozza2010}.

There are numerous software packages that encode the math necessary to model and fit a microlensing event such as 
BAGLE \citep{Lu:2025},
pyLIMA \citep{Bachelet2017}, VBMicrolensing \citep{Bozza2010,Bozza2018,Bozza2021,Bozza2024}, RTModel \citep{BozzaRT:2024} and MulensModel \citep{Poleski2019}.
These packages are all publicly available and have their own strengths and weaknesses in terms of the types of event geometries they support, whether they include both photometry and astrometry in their models, and how accurate and efficient their model-fitting capabilities are. A detailed comparison of these packages for point-source, point-lens models is presented in \citep{Lu:2025}. PyLIMA, VBMicrolensing and MulensModel all support binary lenses, binary sources and both (with orbital motion). Only RTModel (which relies on using VBMicrolensing) and BAGLE support joint-fitting of photometric and astrometric datasets, although RTModel does not yet support joint-fitting for binary-source, binary-lens events. 





In this work, we introduce models for binary systems into the Bayesian Analysis of Gravitational Lensing Events (BAGLE) Python package\footnote{\url{https://github.com/MovingUniverseLab/BAGLE_Microlensing}}. This paper is a companion to \citet{Lu:2025}, which introduces BAGLE and presents models for single-lens object and single-source star events, including point-source, point-lens (PSPL); and finite-source, point-lens (FSPL) models. 
Here we describe the BAGLE implementation of models for point-source, binary-lens (PSBL); binary-source, point-lens (BSPL); and binary-source, binary-lens (BSBL) systems. 
The different binary geometries of microlensing systems are introduced in \S\ref{sec:binary_geometry}.
The general mathematical framework for modeling binary orbital motion in BAGLE is described in \S\ref{sec:binary_solutions}. 
The complete equations of motion are presented for BSPL (\S\ref{sec:binsource}), PSBL (\S\ref{sec:binlenses}), and BSBL  (\S\ref{sec:bineverything}). 
Each of these sections contains sub-sections where we present models for static binaries with fixed primary + secondary positions, secondary companions that move with linear or accelerating motions with respect to their primary, and binary systems with full Keplerian orbital motion. 
Model validation and comparison with other packages is presented in \S\ref{sec:validation}. 
In \S\ref{sec:results}, we present example magnification maps and centroid shift maps from various BAGLE binary models. Changes to photometric lightcurves and astrometric trajectories with binary mass ratio, separation, and orbital parametres are also discussed in this section. 
Conclusions are presented in \S\ref{sec:conclusion} along with planned BAGLE upgrades.

\section{Binary Lens Geometries}
\label{sec:binary_geometry}

\begin{figure}
    \centering
    \includegraphics[width= .5 \textwidth]{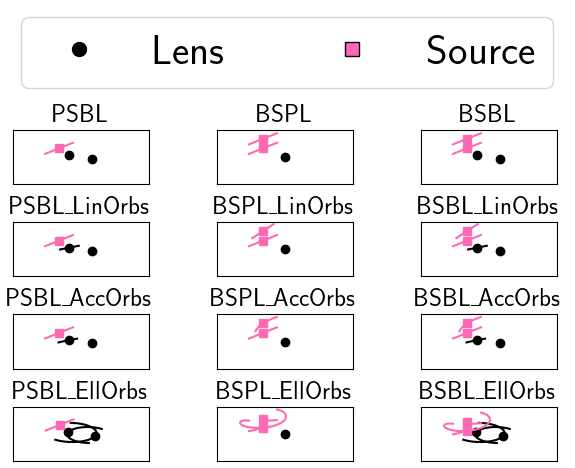}
    \caption{Binary geometries available in BAGLE. Source trajectories on the sky plane are represented in pink and lens trajectories are represented in black. The change in Right Ascension is plotted on the x-axis and the change in Declination is plotted on the y-axis. Models with binary lenses ({\em left column}), binary sources ({\em middle column}), and binary lens and source ({\em right column}) are supported. Secondary companions can have fixed separation and angle relative to the primary ({\em top row}), linear motion ({\em 2nd row}), accelerating motion ({\em 3rd row}), or full Keplerian orbital motion ({\em 4th row}).}
    \label{fig:lens_source_models}
\end{figure}

BAGLE v1.0.1 and later support binary lens and binary source geometries as shown in different columns of Figure \ref{fig:lens_source_models} for PSBL, BSPL, and BSBL systems.
The simplest case of a binary model in BAGLE involves the binary system moving with a fixed angular separation between the primary and secondary objects. This works under the assumption that the period of the orbits is much larger than the duration of the microlensing. For longer-duration events, the secondary companion moves with respect to the primary.

BAGLE provides models for the secondary companion's motion that is linear, accelerating, or orbiting along a Keplerian trajectory (either circular or elliptical) as shown in the different rows of Figure \ref{fig:lens_source_models}. 

\section{Binary Orbital Motion in BAGLE \label{sec:binary_solutions}}

\begin{figure}
    \centering
    \includegraphics[width= 0.5 \textwidth]{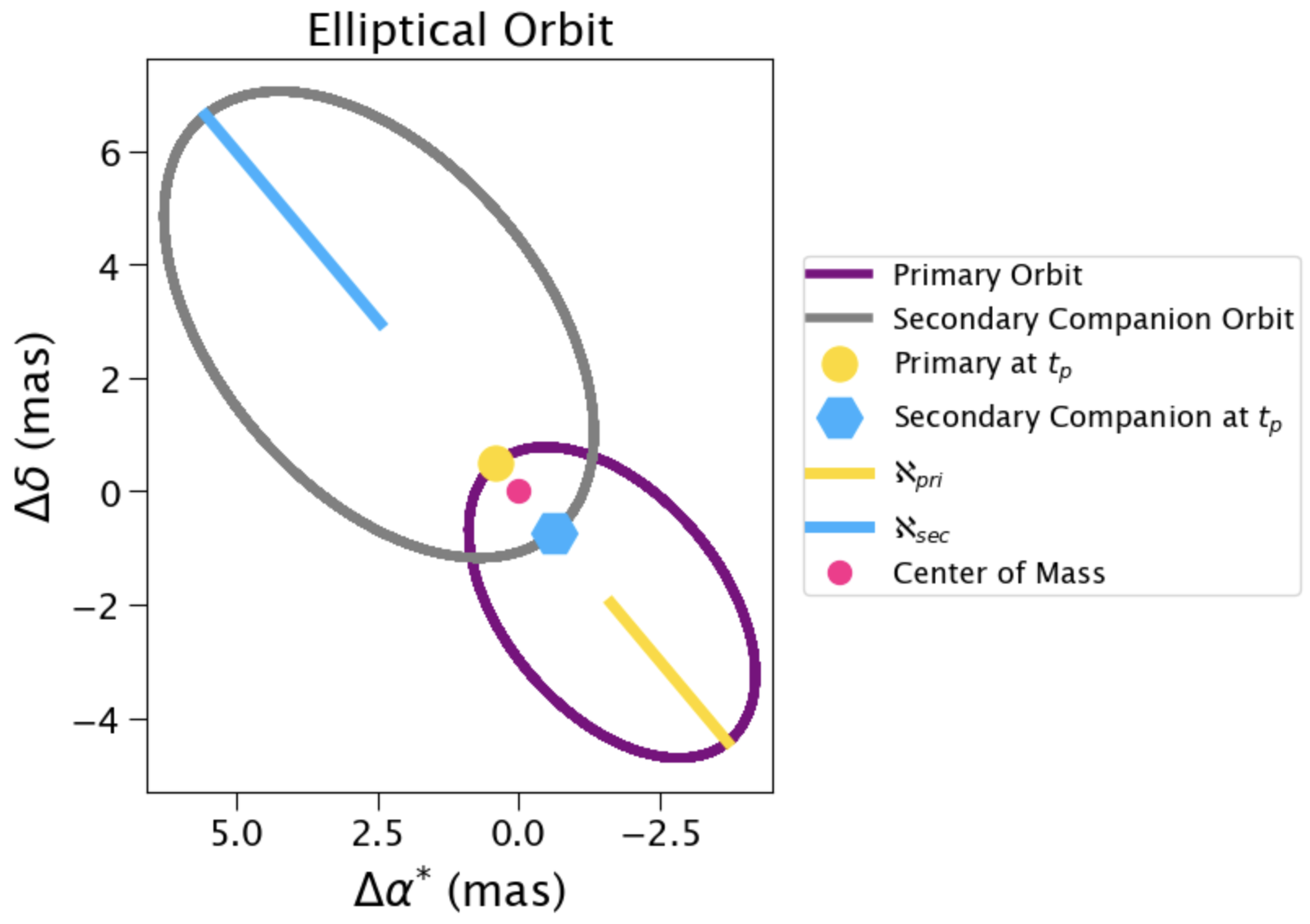}
    \caption{Trajectory of a binary orbit at $\inclination = 0 \degree$ (orbital plane} is face-on) simulated using BAGLE for stationary proper motion. The orbit has an eccentricity of 0.8. The primary and secondary objects at the time of periastron passage have been highlighted.
    \label{fig:com_geometry}
\end{figure}

BAGLE is capable of modeling microlensing events using physical parameters such as the mass of the lens, the distance to the lens, flux, sky position, and proper motion of the lens and the source star. In order to support binary companions to the lens, the source, or both, we introduce many new parameters to describe the mass ratios, flux ratios, and orbital parameters for each binary system. 

To model the primary and secondary motion around the center of mass, the following Keplerian orbital parameters are used: 
\begin{itemize}
\label{keplerian_elems}
    \item $\w$: The argument of periastron of the primary object's orbit in degrees. The secondary companion is placed $180 \degree$ across the primary's argument of periastron. When $\eccentricity = 0$, we set $\w = 0$ since there is no periapsis.
    \item $\inclination$: Inclination angle of the system in degrees. The primary and secondary objects share the same inclination angle.  An inclination of $0\degree$ means that the system is face-on.
    \item $\bigomega$: The longitude of the ascending node of the secondary companion's orbit in degrees. 
    \item $\eccentricity$: Eccentricity of the Keplerian orbit. For circular orbits, this is fixed to 0. 
    \item $\period$: The orbital period of the binary system in days. 
    \item $t_p$: The time of the periastron of the system in days. 
    \item $\al$: The projected semi-major axis of the primary object in mas.
    \item $\ala$: The projected semi-major axis of the secondary object in mas. 
\end{itemize}

Note that not all of the Keplerian elements are required as inputs to BAGLE models. Depending on the nature of the binary object (i.e., whether it is a binary source or a lens) and the parameterization used, the input microlensing parameters vary.
The reference direction is the North direction in the plane-of-sky. Hence, $\bigomega$ is recorded Eastward of North, and $\w$ is Eastward of $\bigomega + 180 \degree$. 

In BAGLE, the eight Keplerian elements presented above are used to estimate the Thiele-Innes constants as described in \citet{Koren_2016}, \citet{Thiele1883} and presented, in detail, in Appendix \ref{sec:Thiele-Innes}. Using these constants, the positions of the primary and the secondary companion over time are computed. Let  $\Xcomp$ be the primary object's projected trajectory (on the sky plane) around the center of mass, with $\Xcomep$ and  $\Xcomnp$ being the East and North components, respectively. Let  $\Xcoms$ be the secondary companion's trajectory around the center of mass with $\Xcomes$ and  $\Xcomns$ representing the East and North components, respectively. Then, the following set of equations holds: 
\begin{eqnarray}
\label{eqn:tinnes}
    \Xcomep &= X(t) \Bpri + Y(t) \Gpri  \nonumber \\
    \Xcomnp &= X(t) \Apri + Y(t) \Fpri  \nonumber \\
    \Xcomes &= X(t) \Bsec + Y(t) \Gsec \nonumber \\
    \Xcomns &= X(t) \Asec + Y(t) \Fsec 
\end{eqnarray}
where $X(t)$ and $Y(t)$ are the rectangular coordinates of the binary system; $A_{pri}$, $A_{sec}$, $B_{pri}$, $B_{sec}$, $G_{pri}$, $G_{sec}$, $F_{pri}$ and $F_{sec}$ are the Thiele-Innes constants. 

Using Eqn~\ref{eqn:tinnes}, BAGLE can model circular and elliptical orbital trajectories around the center of mass. An instance of a Keplerian elliptical orbital trajectory (eccentricity of 0.8) with stationary proper motion is shown in Figure~\ref{fig:com_geometry}. Note that the equations in this section describe the motion around the center of mass. The proper motion of the center of mass in the observer's frame of reference is accounted for separately. 


\section{Binary Sources and Point Lenses (BSPL) \label{sec:binsource}}

In this section, we will discuss microlensing models for binary sources. We begin by discussing simple, static approximations in \S\ref{sec:binsources_static}. Then, we expand to a discussion of linear and accelerated orbital approximations in \S\ref{sec:binsources_lin}. Finally, the full Keplerian solutions are presented in \S\ref{sec:binsources_kep}, followed by a discussion of microlensing equations in \S\ref{sec:binsources_eqn}.

\subsection{Static Approximation}
\label{sec:binsources_static}

In binary source models, the components of the binary (i.e., the primary source and the secondary source companion) are initially at rest relative to each other. On the plane of the sky, in the Solar-System Barycenter (SSB) frame, at $\tnot$, the binary source system initially has a fixed angular separation of:

\begin{equation}
    \vect{s_S}(\tnot) = \Xssvec(\tnot) - \Xspvec(\tnot)
\end{equation}
where $\vect{s_S}(\tnot)$ is the separation vector at $\tnot$, $\Xssvec(\tnot)$ is the initial secondary source position on the sky at $\tnot$, and $\Xspvec (\tnot)$ is the initial primary source position on the sky at $\tnot$. Note, we use the Sun symbol, $\odot$, to indicate SSB coordinates. In the binary source models, $\tnot = \tpnot$, i.e., $\tnot$ is the time of closest approach between the primary source and the lens ($\tpnot$). The angular separation between the primary and the secondary companion is fixed at all times.

\begin{figure*}
    \centering
    \includegraphics[width= \textwidth]{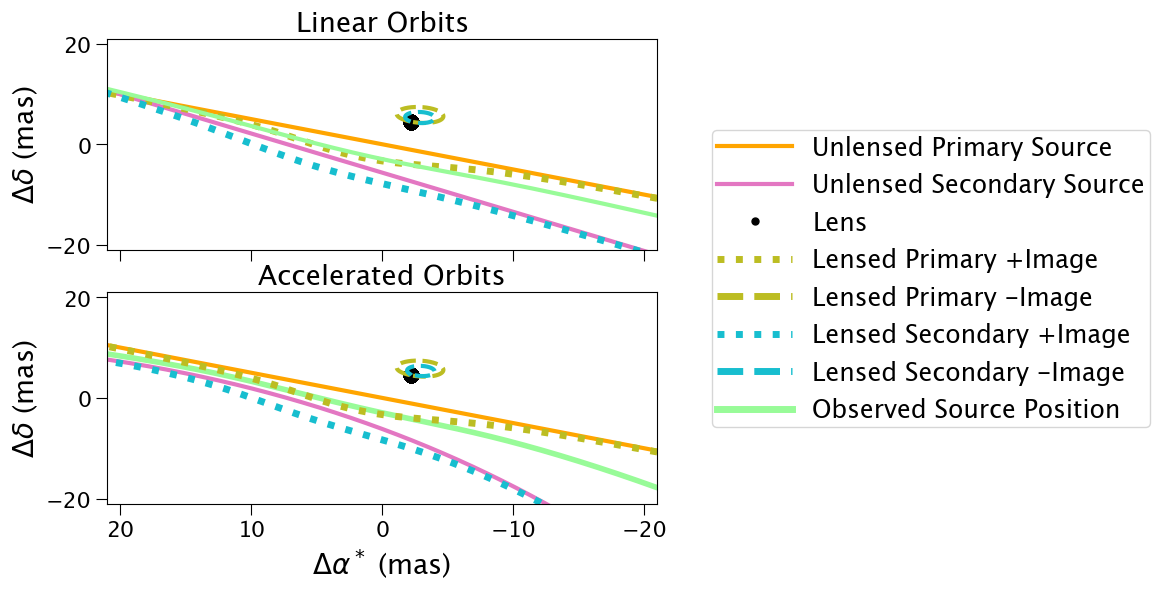}
    \caption{Source and lens trajectories for linear (\emph{Top}) and accelerated (\emph{Bottom}) approximations of orbital motion in binary sources. We present the unlensed sources (solid lines) and the lensed images (dashed + dotted lines). For each source, there is a major and a minor image. The minor image is seen around the lens, and the major image is seen around the source. The green line is the flux-weighted average of lensed source positions, as observed on the sky. Note that $mag_{S,pri}$ = 16 and $mag_{S,sec}$ = 17. In both panels, the primary's proper source motion $\musvec$ is [6 mas yr$^{-1}$, 3 mas yr$^{-1}$]; the secondary's proper source motion relative to the primary source $\mussvec$ is [9 mas yr$^{-1}$, 7 mas yr$^{-1}$]. The acceleration of the secondary source $\accSsec$ is [0.5  mas yr$^{-2}$, -2  mas yr$^{-2}$] in the lower panel. In both cases, the Einstein time $\tE$=269 days, the Einstein radius is 4.9 mas}, and $\uo$ =1.01. The lens mass is $10 M_\odot$, and it is held stationary.
    
    \label{fig:bspl_linacc}
\end{figure*}

\subsection{Linear and Accelerated Orbital Approximations}
\label{sec:binsources_lin}

In contrast to the static approximation, in both the linear and accelerated orbit models, the primary and the secondary companion move with different proper motions. 

For linear approximations, BAGLE inputs a new parameter $\deltamussvec$, which is the proper motion of the secondary source relative to the primary source. With time, due to the proper motion of the primary and secondary sources, the separation vector $\vect{s(t)}$ between the sources changes. The secondary source moves linearly relative to the primary source. The positions of the primary and secondary sources are given by:

\begin{align}
    \Xspvec (t) = & \Xspovec + \musvec (t - \tnot) \nonumber \\
    &+\pi_S \vect{P}(t, \alpha, \delta)  
    \label{linear_motion}    
\end{align}

\begin{align}
    \Xssvec (t) = & \Xssovec + \mussvec (t - \tnot )\nonumber \\
    &+\pi_S \vect{P}(t, \alpha, \delta)  
    \label{linear_motion2}    
\end{align}
where $\musvec$ is the proper motion of the primary source, and $\mussvec$ is the proper motion of the secondary source calculated as $\mussvec = \musvec + \deltamussvec$. Furthermore, we account for the parallactic motion  $\pi_S \vect{P}(t, \alpha, \delta)$ where $\pi_S$ is the parallax amplitude $1/d_S$, and $\vect{P}(t, \alpha, \delta)$ is the actual parallax direction and fractional amplitude on the sky at time $t$ in the direction of the source-lens system $(\alpha, \delta)$, given by the difference of the Earth and Sun's position, normalized by 1 AU. A more complete description of how parallax is defined in BAGLE can be found in \citet{Lu:2025}.

For accelerated approximations, along with $\deltamussvec$, BAGLE also inputs $\accSsec$, which is the acceleration of the secondary source relative to the lens. In the accelerated orbit model, the secondary source moves with constant acceleration relative to the primary source. The positions of the primary and secondary sources are given by: 
\begin{align}
    \Xspvec (t) = & \Xspovec + \musvec (t - \tpnot) \nonumber \\
    &+\pi_S \vect{P}(t, \alpha, \delta)  
    \label{accelerated motion}    
\end{align}

\begin{align}
    \Xssvec (t) = & \Xssovec + \mussvec (t - \tpnot )\nonumber \\
    &+\frac{1}{2}\accSsec[t - \tpnot]^2 +\pi_S \vect{P}(t, \alpha, \delta)  
    \label{accelerated_motion2}    
\end{align}

An example of the linear and accelerated orbital approximations is presented in Figure~\ref{fig:bspl_linacc}, showing the relative change in the projected Right Ascension ($\Delta \alpha^*$) and Declination ($\Delta \delta$) of the primary source (lensed and unlensed), secondary source (lensed and unlensed), and the lens. Here, $\Delta \alpha{^*} = \Delta \alpha \cos{\delta}$. The lensed images are further categorized into the major (``+") image and minor (``-") image. These figures can be generated using BAGLE with the code lines in Table \ref{tab:code_listing_1}. For this example, we use \texttt{BSPL\_PhotAstrom\_noPar\_LinOrbs\_Param2}, which is a model for linear orbits. One can follow the same steps below to generate plots for accelerated orbits.

\begin{table}[ht!]
\centering
\begin{minipage}{0.95\linewidth}
\begin{lstlisting}[language=Python]
from bagle import model

# Define model parameters and pass them into a model instance below.
# mag_base and b_sff are arrays or lists. All other parameters are floats.
# raL and decL must be defined when calling a parallax model.

bsplorbits = model.BSPL_PhotAstrom_noPar_LinOrbs_Param2(
                 t0, u0_amp, tE, thetaE,
                 piS, piE_E, piE_N,
                 xS0_E, xS0_N,
                 muS_E, muS_N,
                 delta_muS_sec_E,
                 delta_muS_sec_N,
                 sep, alpha, fratio_bin,
                 mag_base, b_sff, dmag_Lp_Ls,
                 raL=None, decL=None)

# Define time array
t = np.arange(t0 - 10*tE, t0 + 10*tE, 1)

# Get resolved astrometry for unlensed source positions
xS_unlensed = bsplorbits.get_resolved_source_astrometry_unlensed(t)
srce_pos_primary = xS_unlensed[:, 0, :]
srce_pos_secondary = xS_unlensed[:, 1, :] 

# Get lens astrometry
lens = bsplorbits.get_lens_astrometry(t)

# Get unresolved astrometry for lensed source positions
xS_lensed = bsplorbits.get_astrometry_shift(t)
lensed_pos_pri = xS_lensed[:, 0, :]
lensed_pos_sec = xS_lensed[:, 1, :]
\end{lstlisting}
\caption{Example \bagle code used to initialize the BSPL orbital model and compute unlensed and lensed source astrometry. \label{tab:code_listing_1}}
\end{minipage}
\end{table}

Note that BAGLE has other parameterizations for both linear and accelerated orbits that can be instantiated with different microlensing parameters (e.g., source magnitude instead of baseline magnitude).


\subsection{Full Keplerian Solutions}
\label{sec:binsources_kep}
\begin{figure*}
    \centering
    \includegraphics[width=  \textwidth] {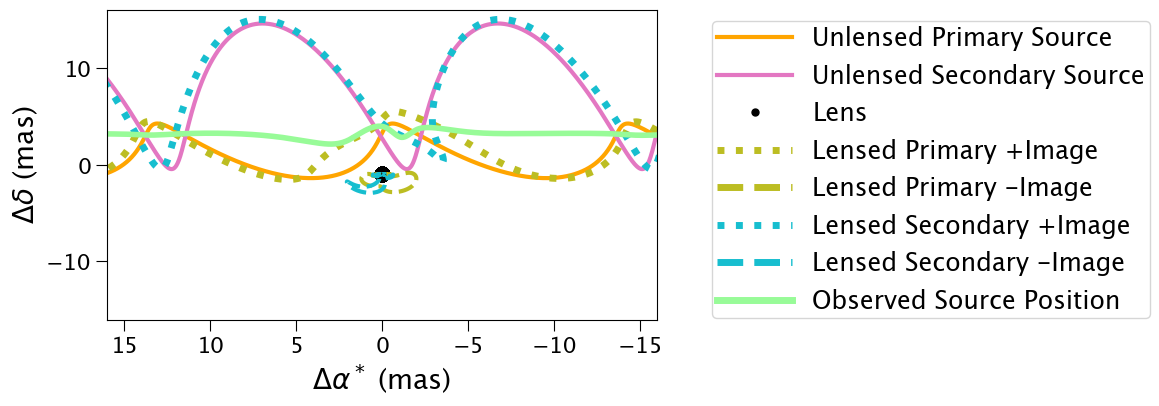}
    \caption{Source and lens trajectories for a simulated binary-source, point-lens microlensing event involving Keplerian orbits. We present the unlensed sources (solid lines) and the lensed images (dashed + dotted lines). The green line is the flux-weighted average of lensed source positions, as observed on the sky. Note that $mag_{S,pri}$ = 16 and $mag_{S,sec}$ = 17. This event has the following Keplerian elements: $\w = 30 \degree$, $\bigomega = 10 \degree$, $\inclination = 90 \degree$, $e=0.6$, $\period = 1000$ days, $\al = 3  $~mas and $\ala = 8 $ mas. The Einstein radius is 2.9 mas. We simulate this event over $t_E=208.47$ days. The lens (8 $M_\odot$) is held stationary.}
    \label{fig:bspl_keplerian}
\end{figure*}

All binary source models with  Keplerian motion in BAGLE input the following Keplerian orbital parameters: $\w$ (set to 0 for circular orbits), $\inclination$, $\bigomega$, $\eccentricity$ (for elliptical orbits), $\period$, $t_p$, $\al$, $\ala$. These models input an additional quantity $\mussysvec$, which is the proper motion of the source system's center of mass in the observer's frame of reference. $\Xscomvec$ is the position of the binary source's center of mass at $\tnot=\tpnot$. While BAGLE inputs $\tnot = \tpnot$ as the time of closest approach between the primary source and the lens, and $\uo$ as the closest approach between the primary source and the lens, it can convert between different coordinate systems (e.g., to use the closest approach between the source system center of mass and the lens). Conversions for
$\tnot$ and $\uo$ are presented in Appendix \ref{sec:u0} and \ref{sec:t0}.





After accounting for the proper motion of the center of mass, the primary and secondary source trajectories (including parallactic motion) are given by: 


\begin{align}
    \Xspvec =& \Xscomvec + \mussysvec  [t - \tpnot] \nonumber \\
    +& \Xcomp + \pi_S \vect{P}(t, \alpha, \delta) \\
    \Xssvec =& \Xscomvec + \mussysvec [t - \tpnot] \nonumber \\
    +& \Xcoms + \pi_S \vect{P}(t, \alpha, \delta) 
\end{align}

We simulate a binary microlensing event to see the effects of complete Keplerian orbital motion in Figure~\ref{fig:bspl_keplerian}. 

\subsection{Lensing a binary source}
\label{sec:binsources_eqn}

The equations of relative separation between each source and lens, in units of Einstein radii, are
\begin{eqnarray}
\upveco &= \frac{\Xspvec-\Xlvec}{\thetaE} \\
\usveco &= \frac{\Xssvec-\Xlvec}{\thetaE}
\end{eqnarray}
in the heliocentric frame.

If the lensing event could be fully resolved, we would expect to see four lensed images, two for each source. 

The magnifications for the images are
\begin{eqnarray}
A_{p,\pm} &= \frac{1}{2} \left(\frac{\upvec^2 + 2}{\upvec \sqrt{\upveco^2 + 4}} \pm1 \right) \\
A_{s,\pm} &= \frac{1}{2} \left(\frac{\usvec^2 + 2}{\usvec \sqrt{\usvec^2 + 4}} \pm 1 \right) 
\end{eqnarray}
where the two images per source are labeled $+$ for the major image and $-$ for the minor image.

Each source's intrinsic flux is magnified by its specific magnification factors. The total magnification for each source is
\begin{eqnarray}
 A_p &= \frac{\upvec^2 + 2}{\upvec \sqrt{\upvec^2 + 4}} \\
 A_s &= \frac{\usvec^2 + 2}{\usvec \sqrt{\usvec^2 + 4}}
\end{eqnarray}

We can define the effective magnification for the system using
\begin{eqnarray}
 A = \frac{A_p F_p + A_s F_s}{F_p + F_s}
\end{eqnarray}
where $F_p$ and $F_s$ are the intrinsic flux of the primary and secondary sources.

The observed flux for the lensed system is then
\begin{eqnarray}
f_{obs} = (A_p F_p + A_s F_s) + (F_p+F_s)\left(\frac{1-\bsff}{\bsff} \right).
\end{eqnarray}
where $\bsff$ is the ratio of the total source flux to the total flux of the source, neighbors, and the lens.  

The image centroid is then simply a flux-weighted combination of the lensed astrometry from the two sources.


\section{Point Sources and Binary Lens (PSBL) 
\label{sec:binlenses}}

In this section, we begin by discussing the static lens approximation in \S\ref{sec:binlenses_static}, followed by the linear and accelerated orbital approximations in \S\ref{sec:binlenses_lin}. In \S\ref{sec:binlenses_kep}, the full Keplerian solutions are presented. Lastly, we discuss the binary lens equations in \S\ref{sec:binlenses_eqn}.

\subsection{Static Approximation}
\label{sec:binlenses_static}

Like binary source models in BAGLE, the simplest implementation of a PSBL model in BAGLE fixes the angular separation between the primary and secondary lens at all times. The two lenses are separated by an initial separation given in Equation~\ref{eqn:sep}.

\begin{equation}
    \label{eqn:sep}
    \vect{s_L}(\tnot) = \Xlsvec(\tnot) - \Xlpvec(\tnot)
\end{equation}

For binary lenses, $\tnot=t_{geom, 0, \sun}$ in static, linear and accelerated orbital approximations. $t_{geom, 0, \sun}$ represents the time of closest approach between the source and the geometric midpoint of the binary lenses. In the static approximation only, the angular separation between the primary and secondary lens is fixed at all times.

\subsection{Linear and Accelerated Orbital Approximations}
\label{sec:binlenses_lin}

\begin{figure*}
    \centering
    \includegraphics[width = \textwidth]{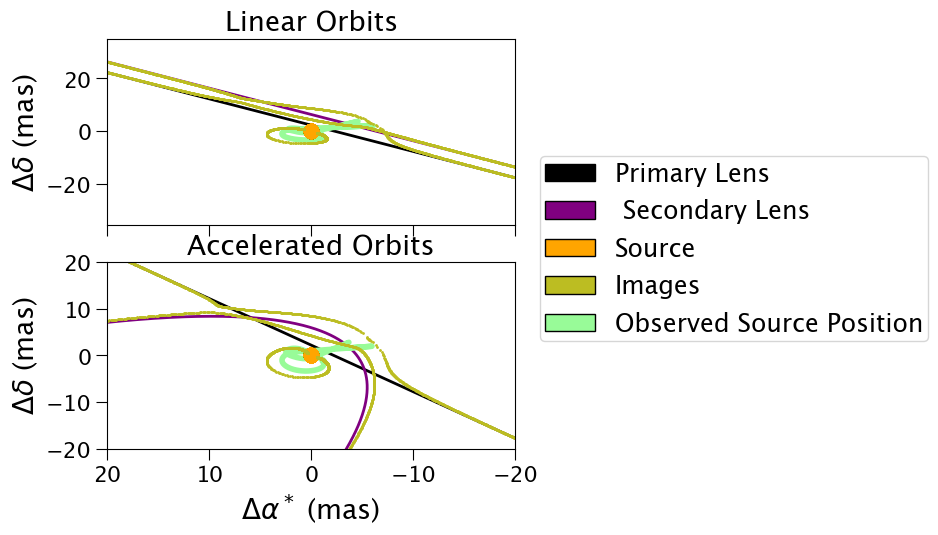}
    \caption{Source and lens trajectories for linear (\textit{upper panel}) and accelerated (\textit{lower panel}) approximations of orbital motion of binary lenses. The solid black line is the primary lens, the solid purple line is the secondary lens, and the solid yellow lines are the image positions. The green line is the flux-weighted average of the lensed source position, as observed on the sky. Note that $mag_{S}$ = 16. In both panels, the source is stationary; the primary lens has a proper motion of $\mulvec$ = [-3.76 mas yr$^{-1}$, -3.76 mas yr$^{-1}$]; the secondary lens has a proper motion of  $\mulsvec$ = [-2.76 mas yr$^{-1}$, -2.76 mas yr$^{-1}$]. For our model with acceleration, we provide the following input for $\accLsec$ = [1 mas yr$^{-1}$, -1 mas yr$^{-1}$]. In both cases, the Einstein time is $\tE$=412 days, the Einstein radius is $6$ mas} and $\uo$ = 0.5. Note that this scenario does not involve a caustic crossing and thus produces only 3 images.
    \label{fig:psbl_linacc}
\end{figure*}




The linear orbital approximations for binary lenses follow the same logic as binary sources in \S\ref{sec:binsources_lin}. Therefore, after accounting for parallactic motion (given by $\pi_L
\vect{P}(t, \alpha, \delta)$ where $\pi_L$ is the parallax amplitude of the lens, $1/d_L$), the positions of the primary and secondary lenses are given by: 

\begin{eqnarray}
\label{linear_motion}
    \Xlpvec (t) =& \Xlpovec + \mulvec [t - t_{geom, 0, \sun}] \nonumber \\
    +& \pi_L \vect{P}(t, \alpha, \delta) \\
\label{linear_motion2}
    \Xlsvec (t) =& \Xlsovec + \mulsvec [t - t_{geom, 0, \sun}]   \nonumber \\
    +&  \pi_L \vect{P}(t, \alpha, \delta) 
\end{eqnarray}
where $\mulvec$ is the proper motion of the primary lens and $\mulsvec$ is the proper motion of the secondary lens. PSBL models with linear orbital approximations input a new parameter $\deltamulsvec = \mulsvec - \mulvec$, which is the proper motion of the secondary lens relative to the primary.

For accelerated approximations, BAGLE takes in $\deltamulsvec$ and another new parameter  ($\accLsec$), which is now the acceleration of the secondary lens relative to the primary lens. The positions of the primary and secondary lenses are given by: 

\begin{align}
\label{accelerated_motion}
    \Xlpvec (t) =& \Xlpovec + \mulvec [t - t_{geom, 0, \sun}] \nonumber \\
    +& \pi_L \vect{P}(t, \alpha, \delta) \\
\label{linear_motion2}
    \Xlsvec (t) =& \Xlsovec + \mulsvec [t - t_{geom, 0, \sun}]   \nonumber \\
    +& \frac{1}{2}\accLsec[t - t_{geom, 0, \sun}]^2 +  \pi_L \vect{P}(t, \alpha, \delta) 
\end{align}

We present examples for linear and accelerated orbital approximations involving binary lenses in Figure~\ref{fig:psbl_linacc}.

\subsection{Full Keplerian Solutions}
\label{sec:binlenses_kep}

\begin{figure*}
    \centering
    \includegraphics[width= \textwidth] {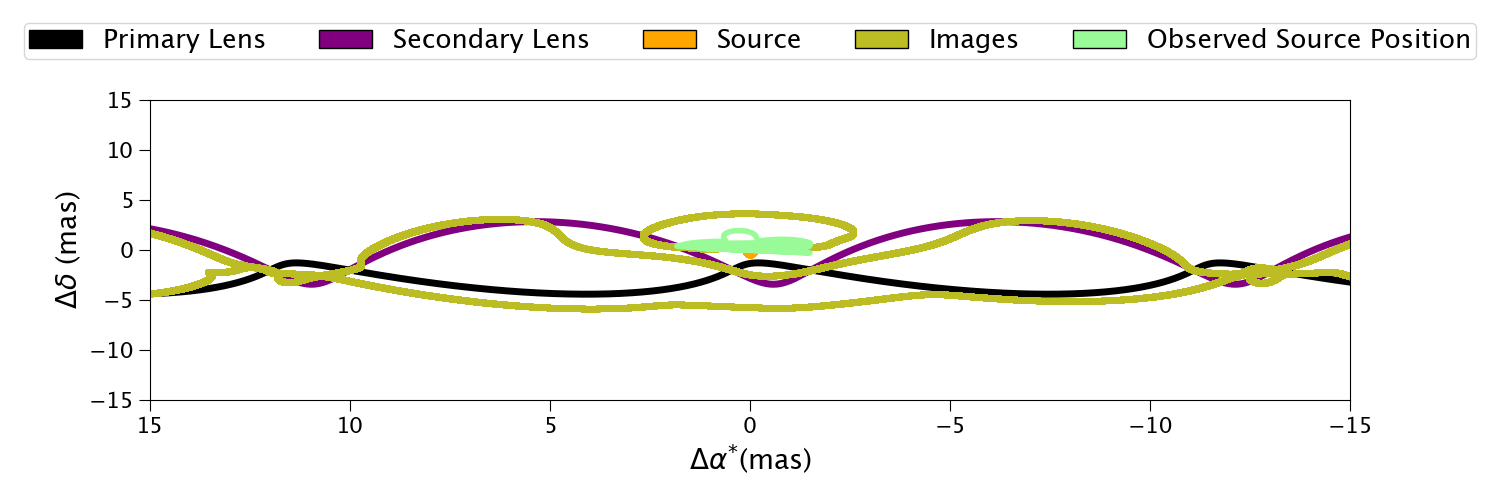}
    \caption{Source and lens trajectories for a simulated microlensing event with a binary lens. The solid black line is the primary lens, the solid purple line is the secondary lens, the orange point is the source position, and the solid yellow lines are the image positions. The green line is the flux-weighted average of the lensed source position, as observed on the sky. Note that $mag_{S}$ = 16. The Einstein radius is $4.5$ mas}. We use the following orbital parameters: $\w = 30 \degree$, $\bigomega = 10 \degree$, $\inclination = 90 \degree$, $e=0.6$, $\period = 1054.41 $ days, and an angular separation of $5$ mas between the two lenses at $\tcomnot$. We present our simulation over $t_E=412.02$ days. The lenses have a mass of $m_{L,p}=$ 10 $M_\odot$ and $m_{L,s}=$ 5 $M_\odot$. The source is stationary.
    \label{fig:psbl_keplerian}
\end{figure*}

All binary-lens models with Keplerian orbital motion (either circular or elliptical) input the following new Keplerian elements: $\w$ (set to 0 for circular orbits), $\bigomega$, $\inclination$, $\eccentricity$, $t_p$, and $a$, where we define $a = \frac{a_{AU}} {d_L}$ such that $a$ is the semi-major axis in units of milli-arcseconds, $a_{AU}$ is the physical semi-major axis in units of AU and $d_L$ is the distance to the lens in units of kpc. In these models, $\mulvec$ is treated as the proper motion of the binary lens system's center of mass instead of the geometric midpoint of the binary lens. As such, we change the notation from $\mulvec$ to $\mulsysvec$ throughout this section. Furthermore, we amend the definition of $\tnot$ for such binary-lens models, i.e, $\tnot = \tcomnot$, where $\tcomnot$ is the time of closest approach between the lens center of mass and the source. The initial position of the lens system's center of mass at $\tcomnot$ is input as $\Xlcomvec$. 


We can use the mass of the primary lens ($m_{L,p}$), the mass of the secondary lens ($m_{L,s}$), and the following set of equations to find the remaining Keplerian elements $\al$,  $\ala$, and $\period$:

\begin{eqnarray}
    \ala &= \frac{m_{L,p}}{m_{L,p}+m_{L,s}} a \nonumber \\
    \al &= a - \ala \nonumber \\
    \period &= 2 \pi \sqrt{\frac{a_{AU}^3}{G(m_{L,p} +m_{L,s})}}
\end{eqnarray}

Using all eight Keplerian elements, the motion of the primary and secondary lenses around their center of mass ($\Xcomp$ for primary and $\Xcoms$ for secondary) at rest can be calculated. Once, the proper motion of the center of mass and the parallactic motion are taken into account, the primary and secondary lens trajectories are given by:

\begin{align}
    \Xlpvec =& \Xlcomvec + \mulsysvec [t - \tcomnot] \nonumber \\
    +& \Xcomp + \pi_L \vect{P}(t, \alpha, \delta) \\
    \Xlsvec =& \Xlcomvec + \mulsysvec [t - \tcomnot] \nonumber \\
    +& \Xcoms + \pi_L \vect{P}(t, \alpha, \delta) 
\end{align}

We can simulate binary lens astrometry trajectories with Keplerian solutions implemented in BAGLE, presented in Figure~\ref{fig:psbl_keplerian}. The code necessary to simulate the astrometric trajectories is displayed in Table \ref{tab:code_listing_2}. We use the PSBL parameterization \texttt{PSBL\_PhotAstrom\_noPar\_EllOrbs\_Param1}, while noting that BAGLE has multiple alternative parameterizations.

\begin{table}[ht!]
\centering
\begin{minipage}{0.95\linewidth}
\begin{lstlisting}[language=Python]
from bagle import model

# Define model parameters and pass them into a model instance below.
# mag_base and b_sff are arrays or lists. All other parameters are floats.
# raL and decL must be defined when calling a parallax model.
#root_tol is set by default to 1e-8. It is a parameter that can be passed to the model instance. 

psblorbits = model.PSBL_PhotAstrom_noPar_EllOrbs_Param1(
            mLp, mLs, t0, xS0_E, xS0_N,
            beta, muL_E, muL_N, omega,
            big_omega, i, e, tp, a, muS_E,
            muS_N, dL, dS, b_sff, mag_src,
            dmag_Lp_Ls, raL=None, decL=None,
            root_tol=1e-8
        )

# Define time array
t = np.arange(t0 - 10*tE, t0 + 10*tE, 1)

# Get resolved astrometry for lenses
lens1, lens2 = psblorbits.get_resolved_lens_astrometry(t)

# Get unlensed source trajectory
source_unlensed = psblorbits.get_astrometry_unlensed(t)

# Get resolved lensed images
images_resolved = psblorbits.get_resolved_astrometry(t)
\end{lstlisting}
\caption{\label{tab:code_listing_2} Example \texttt{bagle} code used to initialize the PSBL orbital model and compute unlensed and lensed source astrometry.}
\end{minipage}
\end{table}


\subsection{Binary Lens Equation}
\label{sec:binlenses_eqn}
The binary lens equation \citep{Schneider_1986} is a mapping of the source position in the ``source plane" to image positions in the ``lens plane", or equivalently ``image plane". The equation is given by
\begin{eqnarray}
    \vect{x}_S = \vect{x}_{obs} - m_1\frac{\vect{x}_{obs} - \vect{x}_{L1}}{|\vect{x}_{obs} - \vect{x}_{L1}|^2} - m_2\frac{\vect{x}_{obs} - \vect{x}_{L2}}{|\vect{x}_{obs} - \vect{x}_{L2}|^2}
\end{eqnarray}
where $\vect{x}_S$ is the angular position of the source (in the source plane), $\vect{x}_{L1}$ and $\vect{x}_{L2}$ are the angular positions of the lenses (in the lens plane), $\vect{x}_{obs}$ is the observed angular position of the lensed images (in the lens plane), and $m_i = \theta_{E, i} ^2 = \frac{4GM_i}{c^2}(\frac{1}{d_L} - \frac{1}{d_S})$, where $M_i$ is the lens mass. 

\begin{figure}
    \centering
    \includegraphics[width=0.45\textwidth]{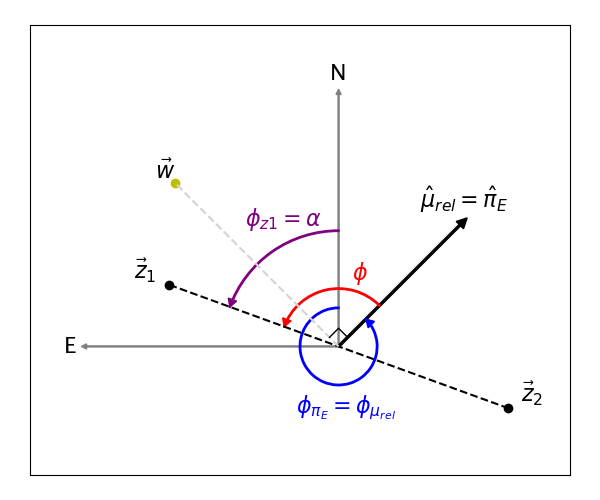}
    \caption{
    Binary lens geometry projected onto the sky in the complex form. 
    The binary lens is at $\vect{z}_1$ and $\vect{z}_2$. The source star 
    is at $\vect{w}$ and the direction of the relative proper motion is shown as $\murelhat$.The angle $\alpha$ is defined as the angle between North and the binary axis. $\alpha$ increments eastwards of North. $\phi_{\pi_E}$ is the angle East of North of $\hat{\mu}_{rel}$ and also $\hat{\pi}_E$. 
    \label{fig:geometry_binary}
    }
\end{figure}

We recast the lens equation in the complex form
\begin{eqnarray}
\label{eqn:lenseqn}
    w = z - m_1 \frac{1}{\bar{z} - \bar{z}_1} - m_2 \frac{1}{\bar{z} - \bar{z}_2}
\end{eqnarray}
where
\begin{eqnarray}
    w &= x_{S,E} + i x_{S,N} \\
    z_1 &= x_{L1,E} + i x_{L1,N} \\
    z_2 &= x_{L2,E} + i x_{L2,N} \\
    z &= x_{obs,E} + i x_{obs,N}
\end{eqnarray}
Figure \ref{fig:geometry_binary} shows how these vectors are projected onto the sky. 

The complex conjugate of Equation ~\ref{eqn:lenseqn} is:
\begin{equation}
    \bar{w} = \bar{z} - m_1 \frac{1}{z - z_1} - m_2 \frac{1}{z - z_2}.
    \label{eqn:binary_zbar}
\end{equation}
The Jacobian, which describes the transformation from source ($w, \bar{w}$) to lens ($z, \bar{z}$) plane is given by
\begin{equation}
    J = 
    \begin{bmatrix}
    \partial w / \partial z & \partial w / \partial \bar{z} \\
    \partial \bar{w} / \partial z & \partial \bar{w} / \partial \bar{z} \\
    \end{bmatrix}.
\end{equation}
Differentiating Eqns. \ref{eqn:lenseqn} and \ref{eqn:binary_zbar} gives
\begin{align}
    \frac{\partial w}{\partial z} &= \frac{\partial \bar{w}}{\partial \bar{z}} = 1 \\
    \frac{\partial \bar{w}}{\partial z} &= m_1 \frac{1}{(z - z_1)^2} + m_2 \frac{1}{(z - z_2)^2}\\
    \frac{\partial w}{\partial \bar{z}} &= m_1 \frac{1}{(\bar{z} - \bar{z}_1)^2} + m_2 \frac{1}{(\bar{z} - \bar{z}_2)^2} = \overline{\frac{\partial \bar{w}}{\partial z}}
\end{align}
which means the determinant of the Jacobian is
\begin{align}
    |J| &= \frac{\partial w}{\partial z} \frac{\partial \bar{w}}{\partial \bar{z}} - \frac{\partial w}{\partial \bar{z}} \frac{\partial \bar{w}}{\partial z} \\
    &= 1 - \left| \frac{\partial \bar{w}}{\partial z} \right|^2.
\end{align}

We can then find the magnification of the source by the lens as:
\begin{equation}
    A = \frac{1}{|J|}.
\end{equation}

There are places in which $|J| \rightarrow 0$. This corresponds to infinite magnification (if the source were a point). The curves in the lens plane where this is true are ``critical curves" and the corresponding curves in the source plane are called ``caustics." In the maps of magnification below (see \S\ref{sec:results_magmap}), we show the source plane in which we can see caustics. We also show the caustics in Figures~\ref{fig:magmaps_varyq} and \ref{fig:magmaps_varysep}. When the source passes over a caustic, this is known as a ``caustic crossing" and changes the number of lensed images of that source from 3 (when the source is outside of the caustic) to 5 (when it is inside the caustic).

To plot the critical curves and caustics, we first solve the equation $0 = |J| = 1 - \left| \frac{\partial \bar{w}}{\partial z} \right|^2$ in the lens plane by noting that the solutions correspond exactly to complex numbers $z$ satisfying 
\begin{align}
\frac{\partial \bar{w}}{\partial z} = m_1 \frac{1}{(z - z_1)^2} + m_2 \frac{1}{(z - z_2)^2} = e^{i\theta}
\end{align}
for some $\theta \in [0,2\pi)$. Clearing the denominators produces a quartic polynomial in $z$ with coefficients depending on $z_1,z_2,m_1,m_2,$ and $\theta$. Points along the critical curve are calculated by solving the four roots of this polynomial for a range of $\theta$ values, and the corresponding caustic curve can then be plotted by using the lens equation to map points from the lens plane to the source plane. Examples of critical and caustic curves are shown in Figure~\ref{fig:magmaps_varyq} and Figure~\ref{fig:magmaps_varysep} for a range of different binary lens separations and mass ratios.

In the microlensing models added to BAGLE, we parameterize our models such that we know where the source is in the source plane ($w$), where the binary lenses are in the lens plane ($z_1$ and $z_2$), the masses of the two lenses, and the distance to all objects. Therefore, we are solving for the observed position of the source in the lens plane ($z$).

In order to avoid working with an equation with a mix of complex numbers and their conjugates, it is standard to plug Eqn.~\ref{eqn:binary_zbar} into Eqn.~\ref{eqn:lenseqn} and simplify \citep[see][]{Witt1990, WittShude1995}. This yields a fifth-order polynomial known as the ``lens polynomial." 

The lens polynomial is solved using \texttt{numpy.roots()} to obtain 5 solutions. The roots are then evaluated in the lens equation and confirmed to satisfy it. There may be some solutions that satisfy the lens polynomial that do not satisfy the lens equation. These are commonly called ``ghost images" or ``ghost roots" \citep[e.g.][]{Bozza2018}. In addition to test the validity of a solution, BAGLE inputs a parameter \textit{root\_tol} for all binary lens models. \textit{root\_tol} is the tolerance threshold in comparing the polynomial roots to the physical solution. The root\_tol defines how close a polynomial root must be to be accepted as a valid solution, and can be changed by the user. By default, it is set to to $10^{-8}$. In cases where the source is far away from the caustic, we  select the correct 3 of the 5 solutions as those that satisfy the lens equation. All valid images contributing to the total magnification are retained, while non-physical roots are rejected using the Jacobian determinant criterion.







\section{Binary Sources and Binary Lenses (BSBL) \label{sec:bineverything}}
 
\begin{figure*}
    \centering
    \includegraphics[width= \textwidth] {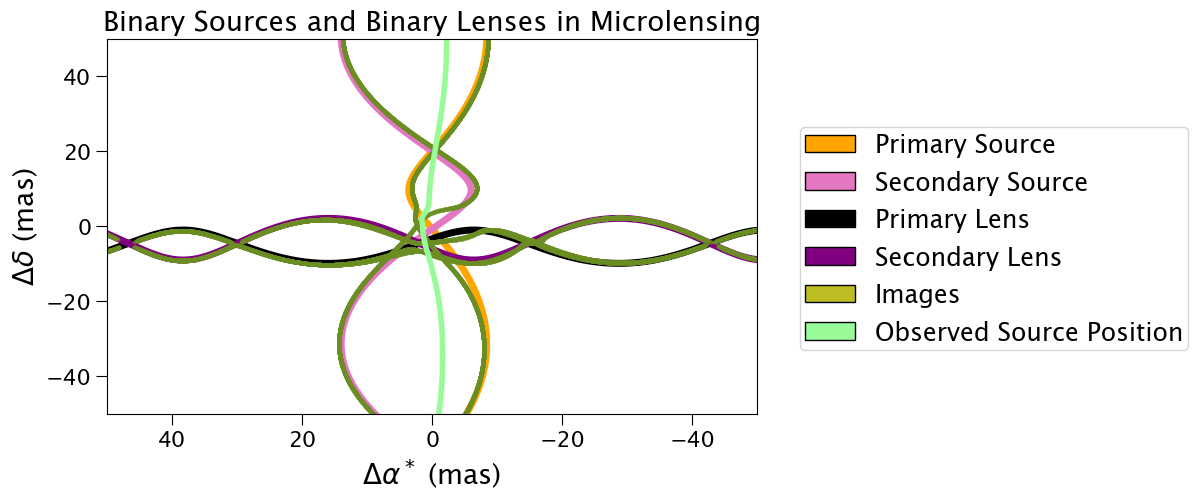}
    \caption{Source and lens trajectories for a simulated binary-source, binary-lens microlensing event with circular orbits. The orange line is the primary source, the pink line is the secondary source, the black line is the primary lens, the purple line is the secondary lens, and the yellow lines are the image positions. The green line is the flux-weighted average of lensed source positions, as observed on the sky. Note that $mag_{S,pri}$ = 16 and $mag_{S,sec}$ = 17. The Einstein radius is $4.9$ mas. The simulation was run with a $t_E = 231.15$ days with the following orbital parameters for the lens: $\w = 30 \degree$, $\bigomega = 10 \degree$, $\inclination = 90 \degree$, $e=0.2$, $\period = 2722.46 $ days, and the following orbital parmaeters for the source $\w = 30 \degree$, $\bigomega = 10 \degree$, $\inclination = 90 \degree$, $e=0.4$, $\period = 6000 $ days. The lenses have a mass of $m_{L,p}=$ 10 $M_\odot$ and $m_{L,s}=$ 8 $M_\odot$.}
    \label{fig:bsbl_keplerian}
\end{figure*}

 BAGLE can also simulate events with both binary sources and binary lenses. The orbital and lensing equations necessary to simulate a binary-source, binary-lens event in BAGLE are individually handled. This means that the equations necessary to simulate binary sources are from \S\ref{sec:binsource}, and the equations necessary to simulate binary lenses are from \S\ref{sec:binlenses}.

For BSBL events with static, linear or accelerated approximations, $\tnot$ is defined as the time of closest approach between the geometric midpoint of the lens and the primary source. However, for Keplerian orbital motion (circular and elliptical), $\tnot$ is defined as the time of closest approach between the center of masses of the binary lens and the binary source. 

We present our astrometric simulations in Figure~\ref{fig:bsbl_keplerian}. \texttt{BSBL\_PhotAstrom\_noPar\_EllOrbs\_Param2} is the BSBL model from BAGLE that is used to simulate an orbital solution for such a microlensing event. The BSBL model assumes that both - the source and the lens - display orbital motion. The figure includes the observed (unresolved) source position on the sky, i.e., the flux-weighted average of lensed source positions. 

\section{Validation of Models}
\label{sec:validation}

\begin{figure}
    \centering
    \includegraphics[width= .48 \textwidth]{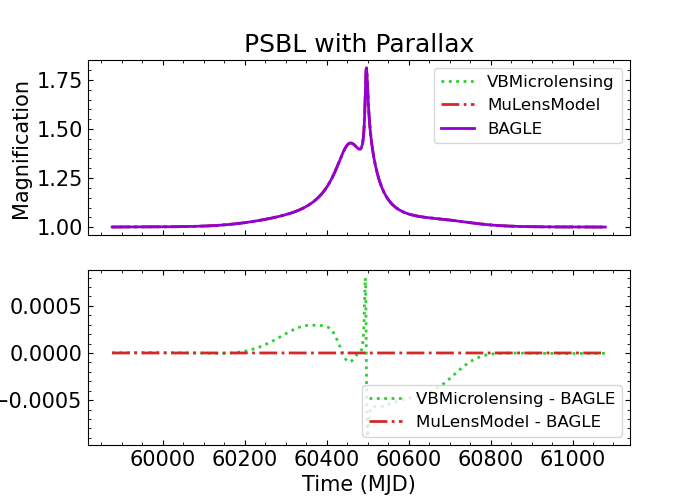}
    \caption{Comparison of a simulated PSBL event with parallax between VBMicrolensing (using \texttt{BinaryLightCurveParallax}), MulensModel and BAGLE. The microlensing parameters in the SSB lens-frame are: $\tnot = 60478.49$ MJD, $\uo= 1 $, $\tE = 100.4$ days, $q = 0.3$, $\piE = [0.3, 0.2]$. The magnification ({\em top}) and residuals with respect to the BAGLE model ({\em bottom}) are shown over time.}
    \label{fig:parallax_comparison}
\end{figure}

\begin{figure}
    \centering
    \includegraphics[width= .48 \textwidth]{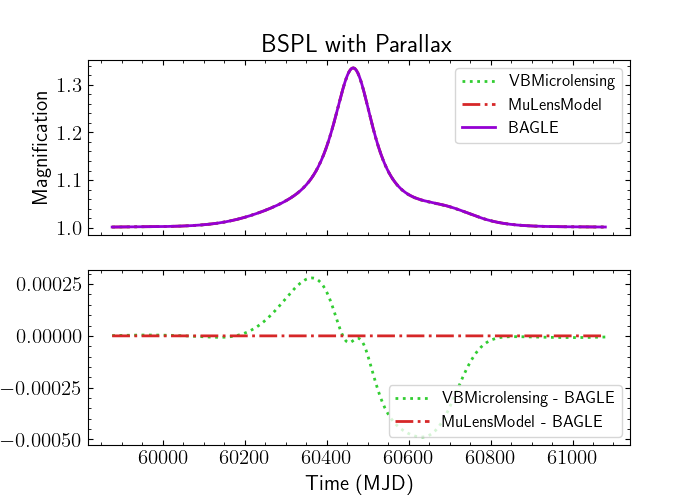}
    \caption{Comparison of a simulated BSPL event with parallax between VBMicrolensing, MulensModel and BAGLE. The microlensing parameters in the SSB lens-frame are: $\tpnot = 60478.49$ MJD, $\tsnot = 60476.09$ MJD, $\upo= 1 $, $\uso= 1.22 $, $\tE = 100.4$ days, $q_{flux, s} = 0.16$, $\piE = [0.3, 0.2]$. The magnification ({\em top}) and residuals with respect to the BAGLE model ({\em bottom}) are shown over time.}
    \label{fig:parallax_comparison_bspl}
\end{figure}

\begin{deluxetable}{lcc}
\tablecaption{PSBL Model-Generation Runtimes for Photometry-Only Models With Parallax
\label{tab:runtimes_par_psbl}}
\tablehead{
  \colhead{} &
  \multicolumn{2}{c}{\textbf{Runtime (ms)}} \\
  \colhead{\textbf{Model}} &
  \colhead{\textbf{Full}} &
  \colhead{\textbf{Pre-Instantiated}}
}
\startdata
BAGLE           &   35.454 $\pm$ 1.183  &  35.171 $\pm$ 0.388  \\
VBMicrolensing  &    3.870 $\pm$ 0.404  &  1.352 $\pm$  0.225 \\
MulensModel     &    4.018 $\pm$ 0.420  &  1.509 $\pm$  0.012  \\
\enddata
\end{deluxetable}

\begin{deluxetable}{lcc}
\tablecaption{BSPL Model-Generation Runtimes for Photometry-Only Models With Parallax
\label{tab:runtimes_par_bspl}}
\tablehead{
   & 
  \multicolumn{2}{c}{{\textbf{Runtime (ms)}}} \\
  \colhead{\textbf{Model}} &
  \colhead{\textbf{Full}} &
  \colhead{\textbf{Pre-Instantiated}}
}
\startdata
BAGLE           &    0.503 $\pm$ 0.029  &  0.451  $\pm$  0.140 \\
VBMicrolensing  &   14.170 $\pm$ 0.460  &   0.318$^{+1.220}_{-0.137}$ \\
MulensModel  &    3.026 $\pm$ 0.415  &   0.617 $\pm$  0.270 
\enddata
\end{deluxetable}

\begin{deluxetable*}{lcccc}
\tabletypesize{\footnotesize}
\tablewidth{0pt}
\tablecaption{Binary Lens and Source Modeling Capabilities Between BAGLE, VBMicrolensing (VBM),  MulensModel (MM) and pyLIMA \label{tab:binary_capabilities}}
\tablehead{
\colhead{Feature} & 
\colhead{BAGLE} &
\colhead{VBM\tablenotemark{*}} & 
\colhead{MM} &
\colhead{pyLIMA}
}
\startdata
\textbf{Static Binary Lenses \& Sources} &
\checkmark &
\checkmark &
\checkmark &
\checkmark \\
\textbf{Complete Binary Lens Orbital Motion} &
\checkmark &
\checkmark &
\checkmark &
\checkmark \\
\textbf{Binary Lens Linear/Accelerated Orbital Approximations} &
\checkmark &
\xmark &
\checkmark &
\xmark \\
\textbf{Complete Binary Source Orbital Motion} &
\checkmark &
\xmark &
\checkmark &
\xmark \\
\textbf{Circular Binary Source Orbital Motion} &
\checkmark &
\checkmark &
\checkmark &
\checkmark \\
\textbf{Binary Source Linear/Accelerated Orbital Approximations} &
\checkmark &
\xmark &
\xmark &
\xmark \\
\multicolumn{5}{l}{\textbf{Photometry Models Available for Fitting}} \\
\textbf{\;\;PSPL} & \checkmark & \checkmark & \checkmark & \checkmark \\
\textbf{\;\;FSPL} & \checkmark & \checkmark & \checkmark & \checkmark \\
\textbf{\;\;BSPL} & \checkmark & \checkmark & \checkmark & \checkmark \\
\textbf{\;\;PSBL} & \checkmark & \checkmark & \checkmark & \checkmark \\
\textbf{\;\;FSBL} & \xmark & \checkmark & \checkmark & \checkmark \\
\textbf{\;\;BSBL} & \checkmark & \checkmark & \checkmark & \checkmark \\
\multicolumn{5}{l}{\textbf{Astrometry Models Available for Fitting}} \\
\textbf{\;\;PSPL} & \checkmark & \checkmark & \xmark & \xmark \\
\textbf{\;\;FSPL} & \checkmark & \checkmark & \xmark & \xmark \\
\textbf{\;\;BSPL} & \checkmark & \checkmark & \xmark & \xmark \\
\textbf{\;\;PSBL} & \checkmark & \checkmark & \xmark & \xmark \\
\textbf{\;\;FSBL} & \xmark & \checkmark & \xmark & \xmark \\
\textbf{\;\;BSBL} & \checkmark & \xmark & \xmark & \xmark \\
\enddata
\tablenotetext{*}{RTModel is used to fit the models generated via VBMicrolensing}
\end{deluxetable*}

In this section, we compare BAGLE with other contemporary microlensing models (VBMicrolensing and MulensModel) against simulated PSBL and BSPL events with parallax, but without orbital motion. Note that we present a comparison for point-source, point-lens models in \citet{Lu:2025}.

A comparison with the inclusion of orbital motion and a comparison with BSBL is reserved for the future. 

We begin by creating a BAGLE model using the parameterization \texttt{PSBL\_Phot\_Par\_Param1}. \texttt{PSBL\_Phot\_Par\_Param1} inputs quantities with reference to the geometric midpoint of the lens system, and calculates Earth's position relative to the Solar System Barycenter over time. Event parameters in BAGLE are: $\tnot = 60478.49$ MJD, $\uo=1$, $\tE = 100.4$ days, $q = 0.3$, $\piE = [0.3, 0.2]$. The event is located in the Galactic Bulge. The relative angle between the binary lens system and the $\murel$ directional vector is $\phi=125 \degree$. The two lenses have a separation of $\vect{s_L}(\tnot) =0.8$. In contrast to BAGLE, other packages prefer a geo-projected frame of reference (as described in \citet{Lu:2025}). The resulting geo-projected parameters are  $\tnotgeotr = 60464.64$ MJD, $\uogeotr= 0.98$, $\tE = 65.06$ days, $q = 0.3$, $\piE = [-0.34, -0.13]$.  After converting to the geo-projected frame of reference, an additional transformation must be applied to $\uvecgeotr$ and $\tnotgeotr$ in order to switch from the geometric midpoint of the lens to the center of mass. The final values input to VBMicrolensing and MulensModel are $\tnotgeotr = 60460.41$ and $\uvecgeotr = 0.78$. In the geo-projected frame, the heliocentric rectilinear frame has the same position and velocity as the true geocentric non-rectilinear frame at a reference time, denoted as $t_{0, par}$ in VBMicrolensing and MulensModel. For simplicity, we let $t_{0, par} = \tnot = 60478.49$ MJD in the comparison.

The comparison between VBMicrolensing, MulensModel, and BAGLE for PSBL events is shown in Figure~\ref{fig:parallax_comparison}. pyLIMA was not included because it uses VBMicrolensing as a backend. The difference in light curves between MulensModel and BAGLE is extremely small around ~10$^{-11}$. In contrast, differences of order 10$^{-4}$ are apparent between BAGLE and VBMicrolensing. It is unclear why the difference between VBMicrolensing and BAGLE exists, and our future work involves resolving it.



Next, a BSPL event pointing towards the Galactic Bulge was compared between VBMicrolensing, MulensModel and BAGLE. The parameterization \texttt{BSPL\_Phot\_Par\_Param1} was used to create a BAGLE model with the following microlensing parameters in the SSB reference frame: $\tpnot = 60478.49$ MJD, $\tsnot = 60476.09$ MJD, $\upo=1$, $\uso=1.25$, $\tE = 100.4$ days, $q_{flux, s} = 0.16$, $\piE = [0.3, 0.2]$; where $\uso$ is the closest approach between the secondary source and lens, $\tsnot$ is the time of this closest approach, and $q_{flux, s}$ is the ratio between the secondary source's flux and the primary source's flux. In the geo-projected frame, $\tpnotgeotr = 60464.64$ MJD, $\tsnotgeotr = 60461.01$ MJD, $\upogeotr=0.98$,  $\usogeotr=1.22$, $\tEgeotr = 65.06$ days, $\piEvecgeotr = [-0.34, -0.13]$. The comparison between different BSPL models is presented in Figure~\ref{fig:parallax_comparison_bspl}. Differences of 10$^{-4}$ are apparent between BAGLE and VBMicrolensing, likely due to differences in the parallax implementations. BAGLE and MulensModel have a much smaller difference around 10$^{-15}$. This level of difference is consistent with the point-source, point-lens comparison (with parallax) presented in \citet{Lu:2025}.

Next, we compare the computational run time of \bagle~with VBMicrolensing and MulensModel. These runtimes are based on the time taken to model the events presented in Figures~\ref{fig:parallax_comparison} and \ref{fig:parallax_comparison_bspl}.
Runtimes are calculated for models by generating mock data for an event with $t_E=100$ days sampling 2000 time steps over 5.5 years. The time tests are repeated 100 times and the mean and standard deviation are recorded for test.
Tests were performed with \bagle~1.0.5, VBMicrolensing v5.3.6, and MulensModel v3.3.1. A 2021 iMax equipped with a M1 processor using a Python 3.11 environment and the ipython kernel (Python gcc $=$ 14.0.0) was used to conduct these tests. A summary of model-generation runtimes is presented in Table~\ref{tab:runtimes_par_psbl} and \ref{tab:runtimes_par_bspl} for PSBL and BSPL events, respectively. 

The pre-instantiated runtime is applicable to running many different sets of model parameters for an event at the same sky location (i.e.~model fitting). On the other hand, the full runtime is applicable to large-scale simulations of multiple events located at different sky coordinates. For pre-instantiated runtimes, we find that all packages are quick for BSPL events. BAGLE is the quickest for BSPL model generations when calculating the full runtime. However, BAGLE is the slowest for PSBL model generation (pre-instantiated and full). Future work involves implementing JAX \citep{jax2018github} to improve the efficiency of the root solver algorithm in BAGLE.


Lastly, a summary of binary modeling capabilities between different packages is provided in Table \ref{tab:binary_capabilities}. pyLIMA is included in this summary. All four packages support static binary lenses and sources. In terms of orbital motion of binary lenses, all packages can simulate elliptical or circular Keplerian orbits that are close to the actual solutions. BAGLE can model linear and accelerated approximations, which are computationally resourceful in microlensing events with very long orbital periods. MulensModel also supports simpler approximations with the inputs $\frac{ds}{dt}$ (rate of change of binary lens separation) and $\frac{d\alpha}{dt}$ (rate of change of the angle between the binary axis and the proper motion vector). In terms of orbital motion for binary sources (often referred to as ``xallarap"), all packages support circular orbital motion. MulensModel can handle elliptical orbital motion and BAGLE provides a linear and accelerated approximation too along with full elliptical orbital motion. It is important to re-iterate that pyLIMA and MulensModel implement a photometry-only fitting using their binary lens and source models. RTModel (developed to use VBMicrolensing for fitting events) can support joint photometric and astrometric fitting for all binary models (including finite sources) except binary-source, binary-lens models. BAGLE can support joint photometric and astrometric fitting of all binary models without finite sources. Note that BAGLE still provides support for finite source effects with point lenses.

\section{Results}
\label{sec:results}

The inclusion of binary models in the \bagle~package enables us to explore many different aspects of binary microlensing events. In this section, we explore some of the most notable impact of including binary models in the \bagle~package.

In this section, we investigate
\begin{itemize}
    \setlength\itemsep{0em}
    \item \S\ref{sec:results_magmap} - Magnification Maps. 
    \item \S\ref{sec:results_asm} - Centroid Shift Maps. 
    \item \S\ref{sec:results_q} - Dependency of Caustics on Mass Ratio and Separation.
    \item \S\ref{sec:results_om} - Dependency of Fitting on Orbital Motion.
\end{itemize}

\subsection{Results: Magnification Maps}

\begin{figure}
    \centering
    \includegraphics[width= 0.48 \textwidth]{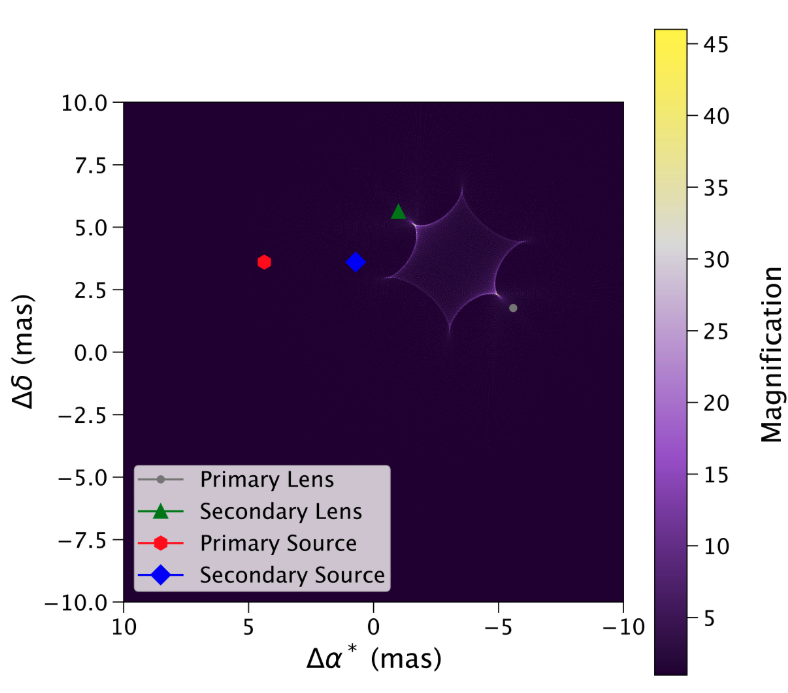}
    \caption{Magnification map for a BSBL microlensing event with both the source and lens. Regions higher on the color scale indicate regions with higher magnification. The magnification map produced in \bagle~for a BSBL event only depends on the lens geometry. The secondary source in the map contributes through flux-weighting. Therefore, this magnification map would be the same for an equivalent PSBL event with the same lens geometry.}
    \label{fig:magmaps}
\end{figure}

\label{sec:results_magmap}

By solving the lens equation, the magnification of the source at any (projected) position relative to the position of the lens can be calculated. The magnification map is a visualization reflecting the calculated magnification at a given point. Magnification maps are generated using the inverse ray shooting method \citet{Bennett_2010}, which inverts the lens equation by shooting rays backward from the observer to the lens plane. 

To make a magnification map in BAGLE, these are the steps we follow:

\begin{itemize}
    \item Create a grid of image positions in the lens frame of reference. 
    \item By mapping the image grid to the source plane and then binning the resulting source positions, we produce a map of ray density, i.e., a map of the number of rays that land at each source position.
    \item The source plane is pixelated, and the number of rays per pixel is proportional to the magnification of a source located at that pixel. The ray density map is normalized by the mean unlensed ray density to produce a magnification map.

\end{itemize}

The magnification map for a BSBL event is presented in Figure ~\ref{fig:magmaps}. The event has a lens mass ratio $q=1$ and a separation of $6$ mas between the lens at $t_p$. The Einstein radius is $4.9$ mas. At $t_p$, the sources have a separation of $13$ mas. The primary source has a magnitude of $16$, the secondary source has a magnitude of $20$ and $\bsff=1$. Note that the secondary source contributes through flux-weighting, which is why a singular caustic structure is visible on the map in Figure~\ref{fig:magmaps}.

A magnification map can be generated using a call to the \texttt{get\_magnification\_map} function in \texttt{plot\_models.py}. The function is called on a model instance, where the complete source and lens trajectory are specified. The \texttt{get\_magnification\_map} function can also be used on a PSBL model instance.

\subsection{Centroid Shift Maps}
\label{sec:results_asm}

\begin{figure}
    \centering
    \includegraphics[width= 0.48 \textwidth]{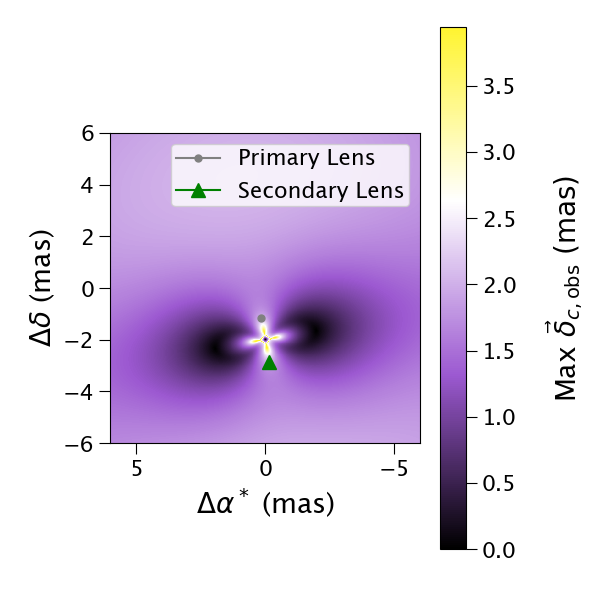}
    \caption{A color map for the centroid shift during a PSBL microlensing event. The color scale indicates the absolute centroid shift ($\vect{\delta_{c, obs}}$) in the image plane.}
    \label{fig:csmap}
\end{figure}

Like the magnification map, a forward ray shooting method can be used to shoot from the source plane to the image plane, creating a color map of the absolute centroid shift. This forward ray shooting method works in the following way:

\begin{itemize}
    \item Create a sample grid of source positions. 
    \item Solve the lens equation to get all possible image positions using the sample grid of source positions. Find the flux-weighted centroids of these possible image positions.
    \item Bin the flux-weighted centroids to create a color map of how many rays fall into each pixel in the image plane grid. Normalize it by using the flux-weighted centroids on the colorbar. 
\end{itemize}

A centroid shift color map for a PSBL event is presented in Figure~\ref{fig:csmap}. The event has a lens mass ratio $q=1$ and a separation of $5$ mas between the lenses at $t_p$. The Einstein radius is $5.2$ mas. The source has a magnitude of 20 and $\bsff=1$.

A centroid shift map can be generated using a call to the \texttt{get\_centroid\_shift\_map} function in \texttt{plot\_models.py}. The function performs the forward-ray-shooting method and plots a centroid shift map. Like magnification maps, an arbitrary time can be provided as an argument when calling the function.

\subsection{Results: Dependency of Caustics on Mass Ratio and Separation}
\label{sec:results_q}

\begin{figure*}
    \centering
    \includegraphics[width= \textwidth]{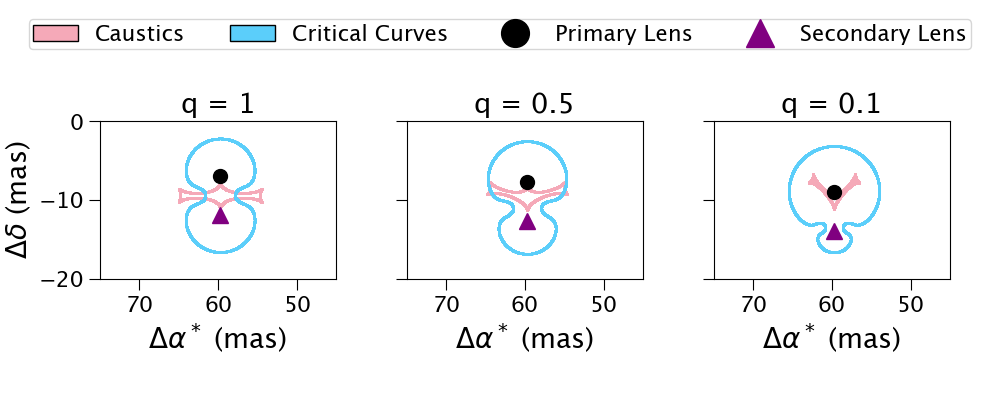}
    \caption{Caustics (\emph{pink}) and critical curves (\emph{blue}) for a PSBL model with the lenses at a separation of 5 mas with $\thetaE = 6$ mas. We present panels with $q$ ranging from 0.1 to 1.}
    \label{fig:magmaps_varyq}
\end{figure*}

\begin{figure*}
    \centering
    \includegraphics[width= \textwidth]{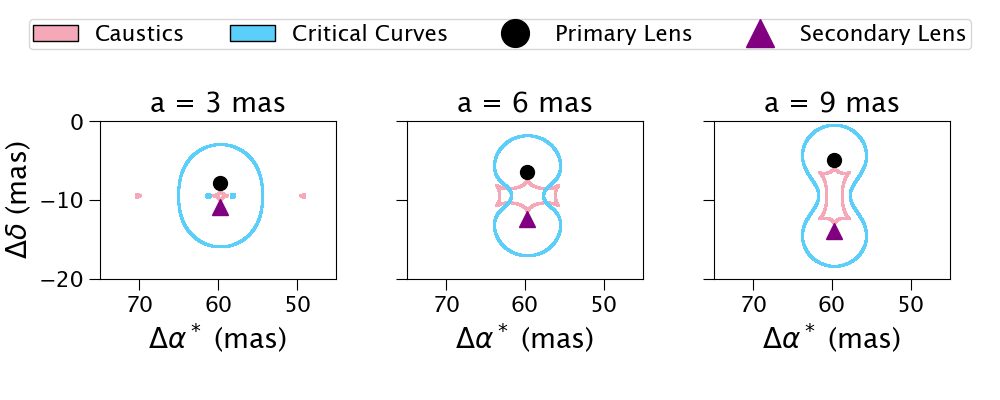}
    \caption{Caustics (\emph{pink}) and critical curves (\emph{blue}) for a PSBL model at $t_p$ with a fixed $q=1$ with $\thetaE = 6$ mas. The value of $a$ (provided as an input to the PSBL models) varies between $3$ mas, $6$ mas, $9$ mas in the three panels.}
    \label{fig:magmaps_varysep}
\end{figure*}

In this section, we explore the dependency of caustics on mass ratios and separation. The separation is fixed at an arbitrarily chosen value of 5 mas and the mass ratio $q=\frac{m_{L,s}}{m_{L,p}}$ for a dark, non-planetary lens, ranges from 0.1 to 1. The Einstein radius is 6 mas. The caustics and critical curves are presented in Figure~\ref{fig:magmaps_varyq}. We see that the caustic becomes more symmetric as $q \rightarrow 1$. The degree of asymmetry is directly dependent on $q$  \citep{Griest1998}.

Similarly, by fixing the mass  $q=\frac{m_{L,s}}{m_{L,p}} = 1$, BAGLE can simulate a dark, non-planetary lens at varying angular separations between the primary and secondary lens. The Einstein radius is again 6 mas. From the caustics and critical curves presented in Figure~\ref{fig:magmaps_varysep}, we see that the caustic remains symmetric regardless of separation for a fixed $q=1$. At larger separations, the caustics also become increasingly elongated or stretched. Detailed descriptions of the geometry of caustics for the q = 1 case can be found in \cite{Schneider_1986}.  

In BAGLE, critical curves and caustics can be created by calling upon the \texttt{get\_critical\_curves} and \texttt{get\_caustics} functions in \texttt{plot\_models.py}.







   %


\begin{figure*}
    \centering
    \includegraphics[width= .32 \textwidth]{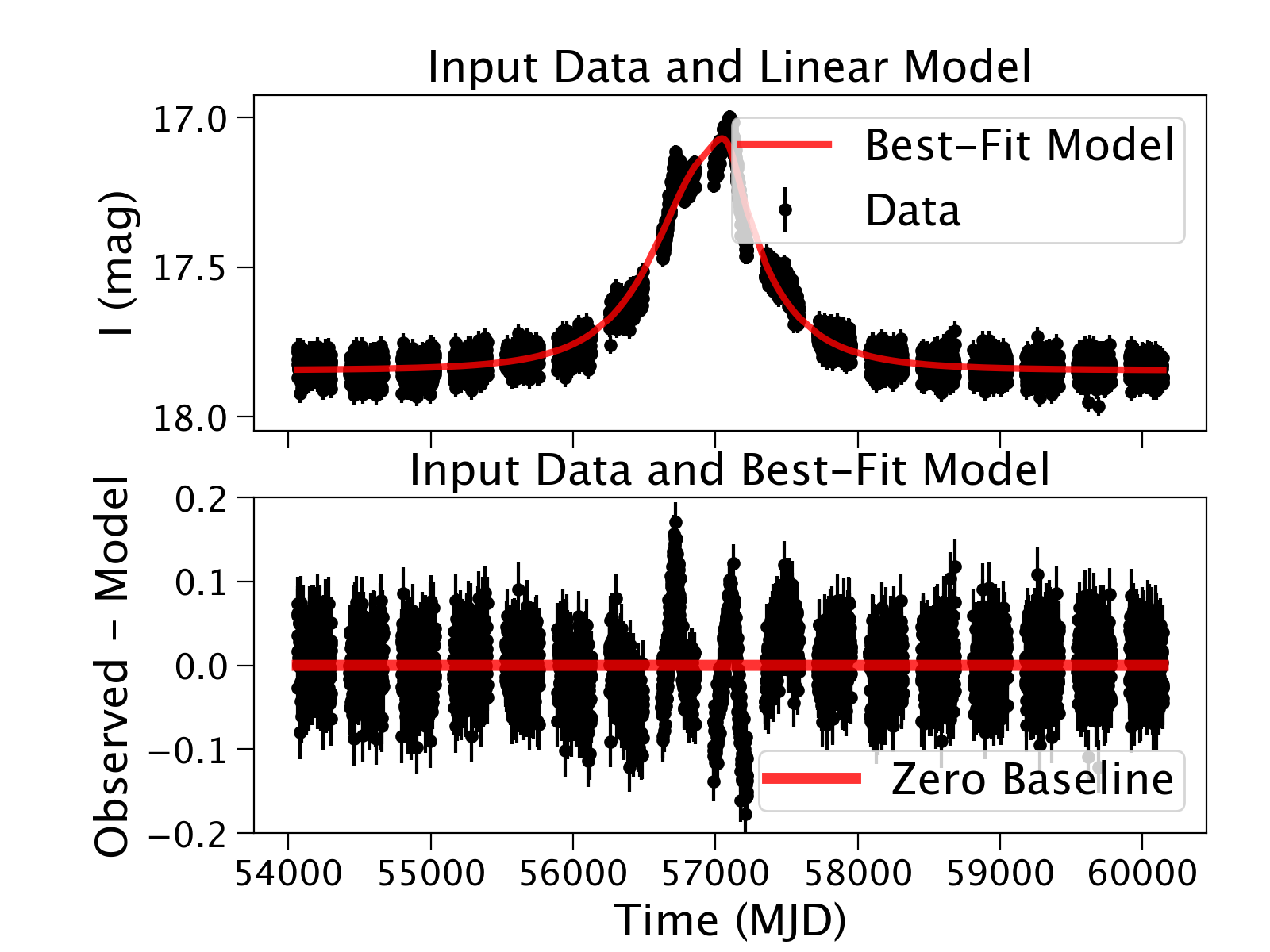}
    \includegraphics[width= .32\textwidth]{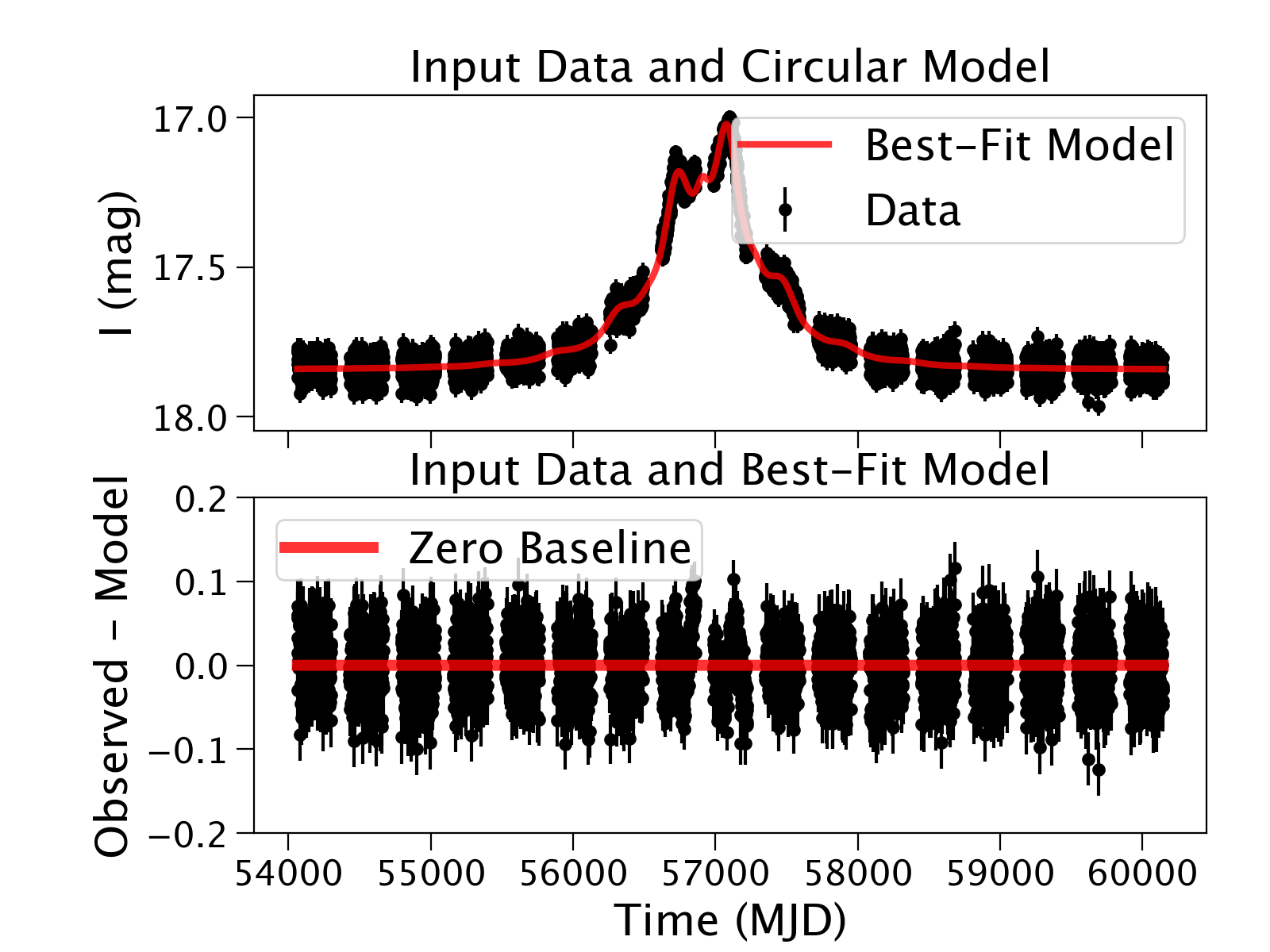}
    \includegraphics[width= .32\textwidth]{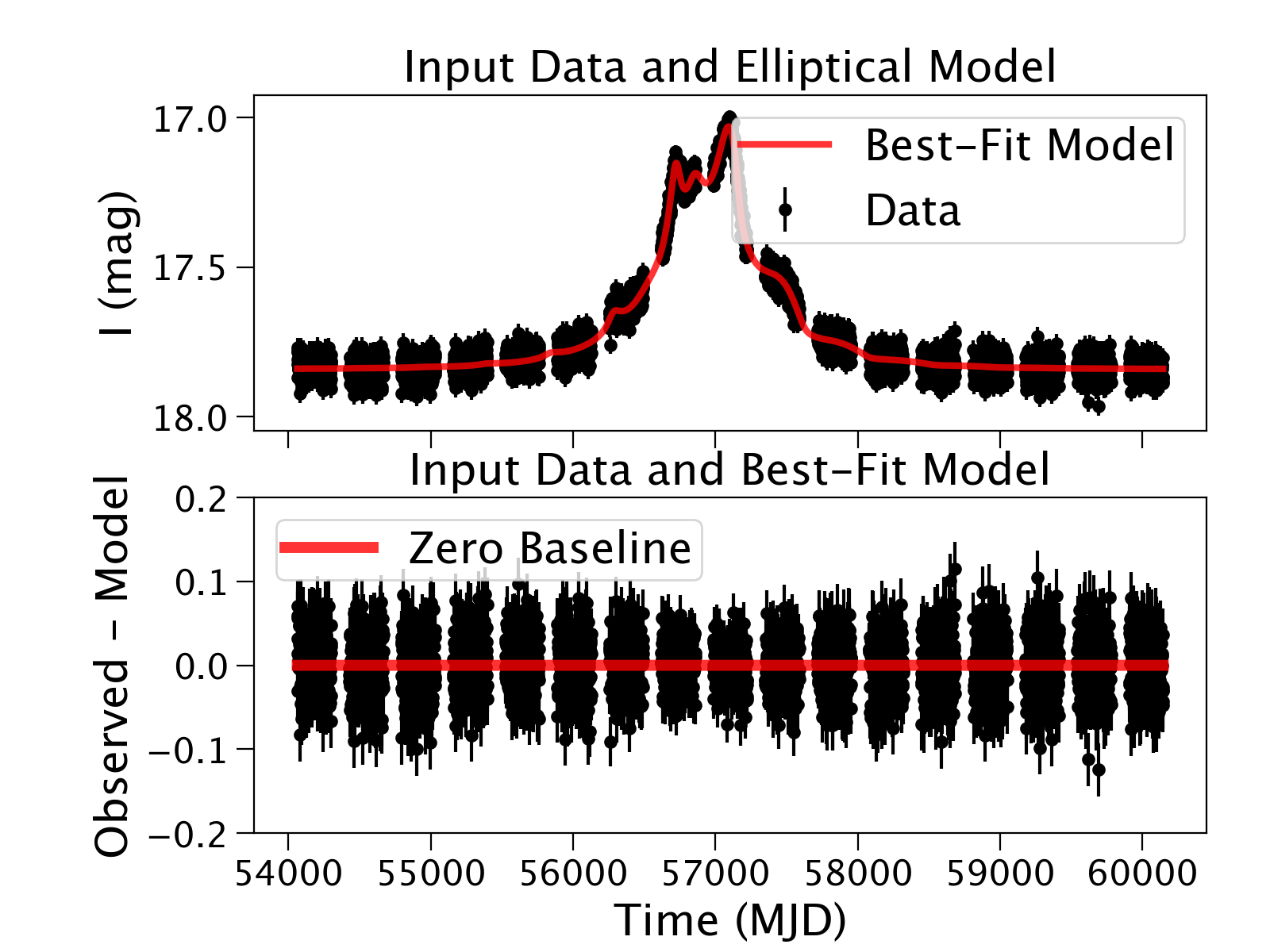}

    \caption{Fitting output for a mock photometric dataset generated using  BSPL\_PhotAstrom\_noPar\_EllOrbs\_Param1 and the parameters displayed in Table~\ref{tab:fake_fit}. We see correlated residuals in the linear fit, unlike the circular and elliptical fits. 
    (\emph{Left}) Best-fit with linear orbital motion. (\emph{Center}) Best-fit with circular orbital motion. (\emph{Right}) Best-fit with elliptical orbital motion.}
    \label{fig:orbital_comparison}
\end{figure*}

\begin{figure*}
    \centering
    \includegraphics[width= .32 \textwidth]{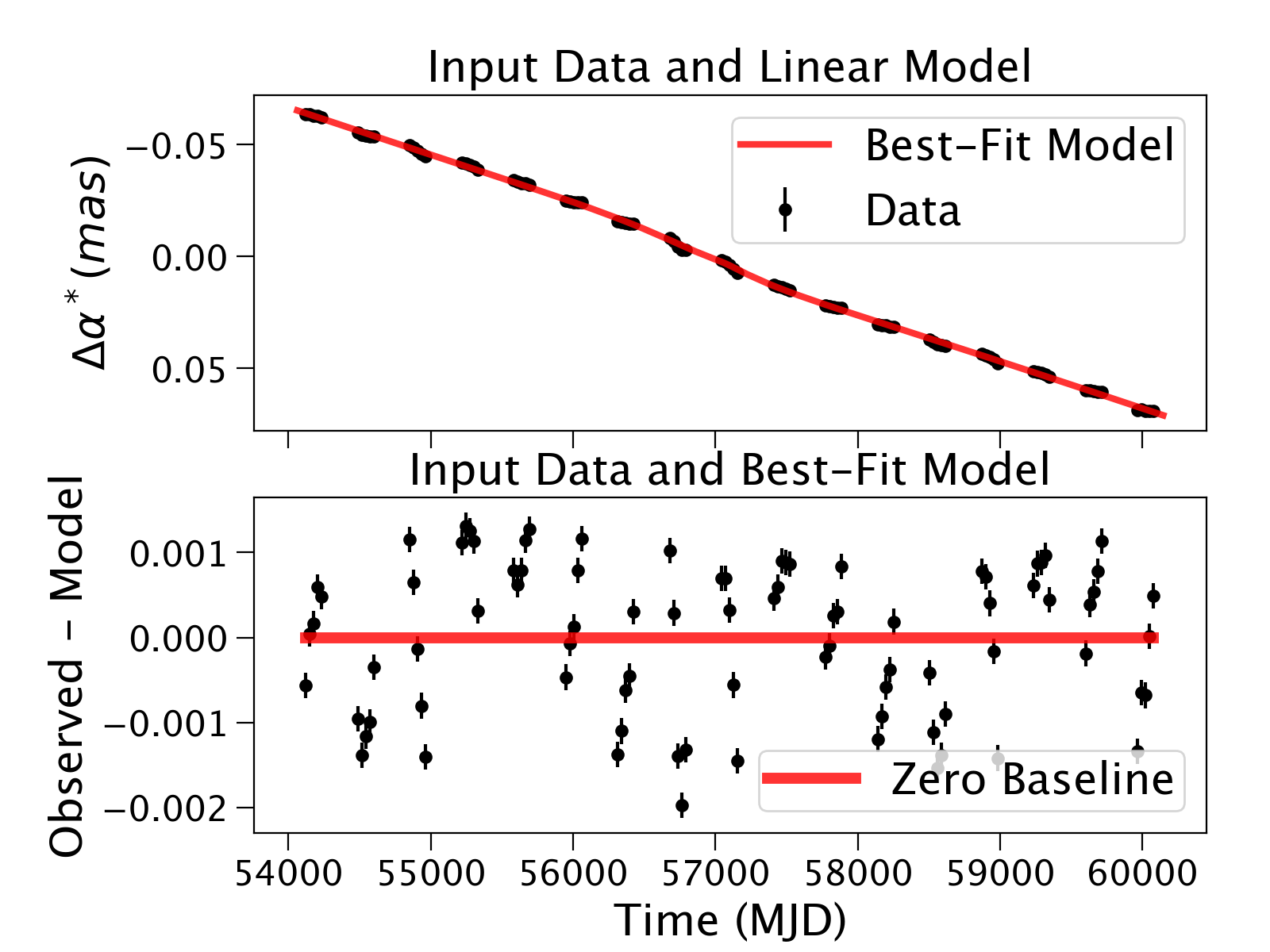}
    \includegraphics[width= .32\textwidth]{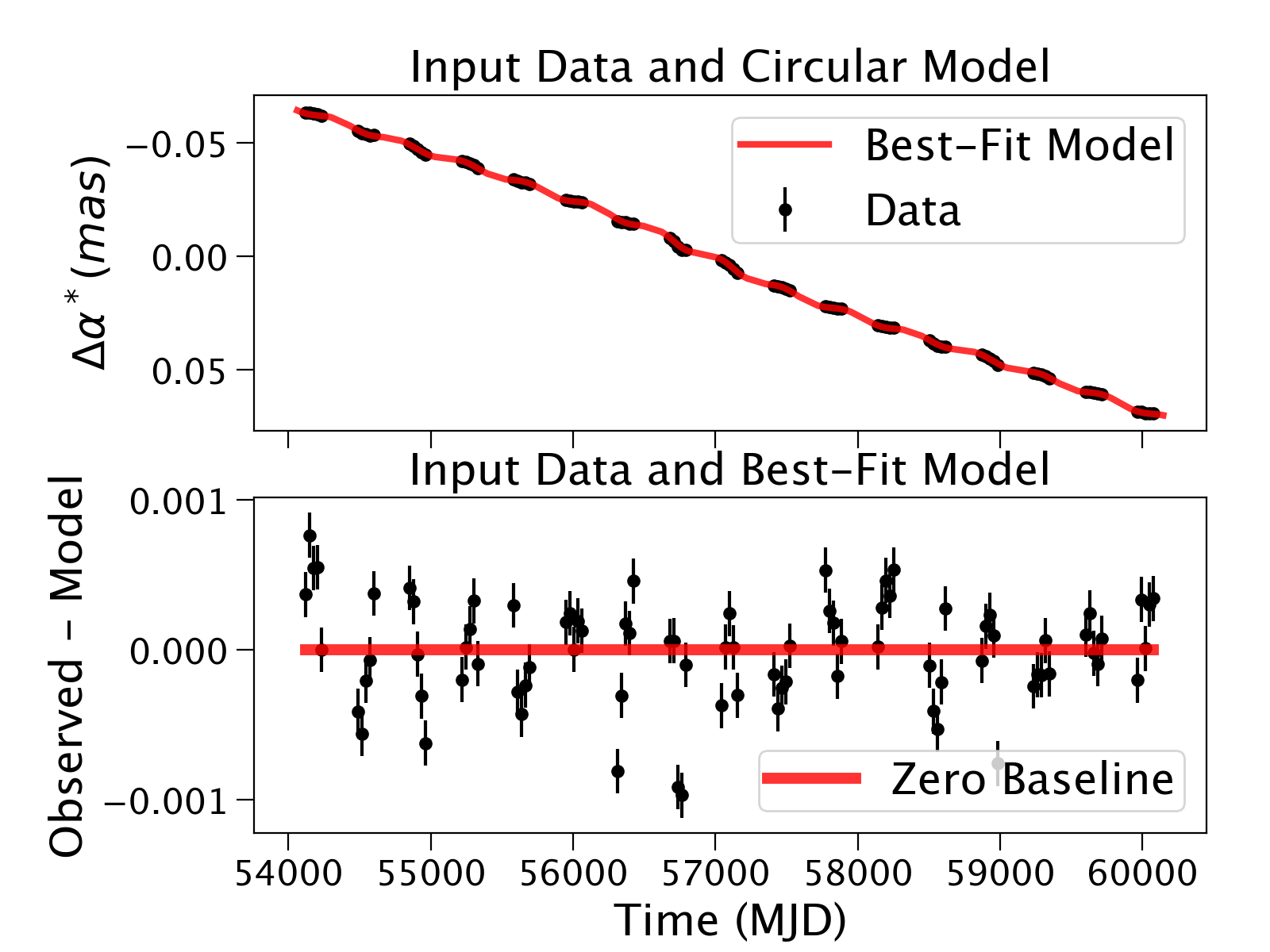}
    \includegraphics[width= .32\textwidth]{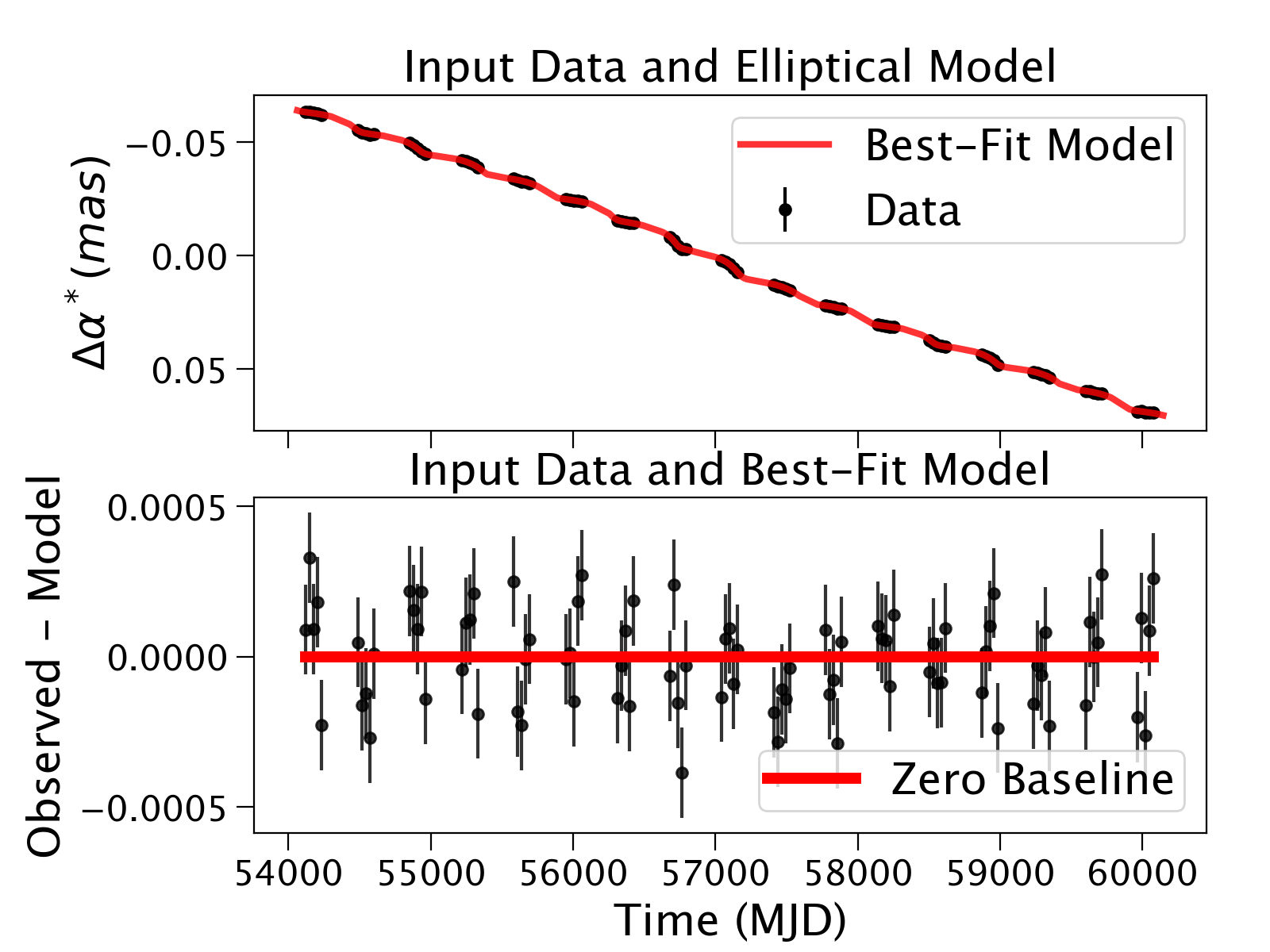}

    \caption{Fitting output for a mock astrometric dataset generated using  BSPL\_PhotAstrom\_noPar\_EllOrbs\_Param1 and the parameters displayed in Table~\ref{tab:fake_fit}. We present a fit for the RA component in this figure. 
    (\emph{Left}) Best-fit with linear orbital motion. (\emph{Center}) Best-fit with circular orbital motion. (\emph{Right}) Best-fit with elliptical orbital motion.}
    \label{fig:orbital_comparison_as1}
\end{figure*}

\begin{figure*}
    \centering
    \includegraphics[width= .32 \textwidth]{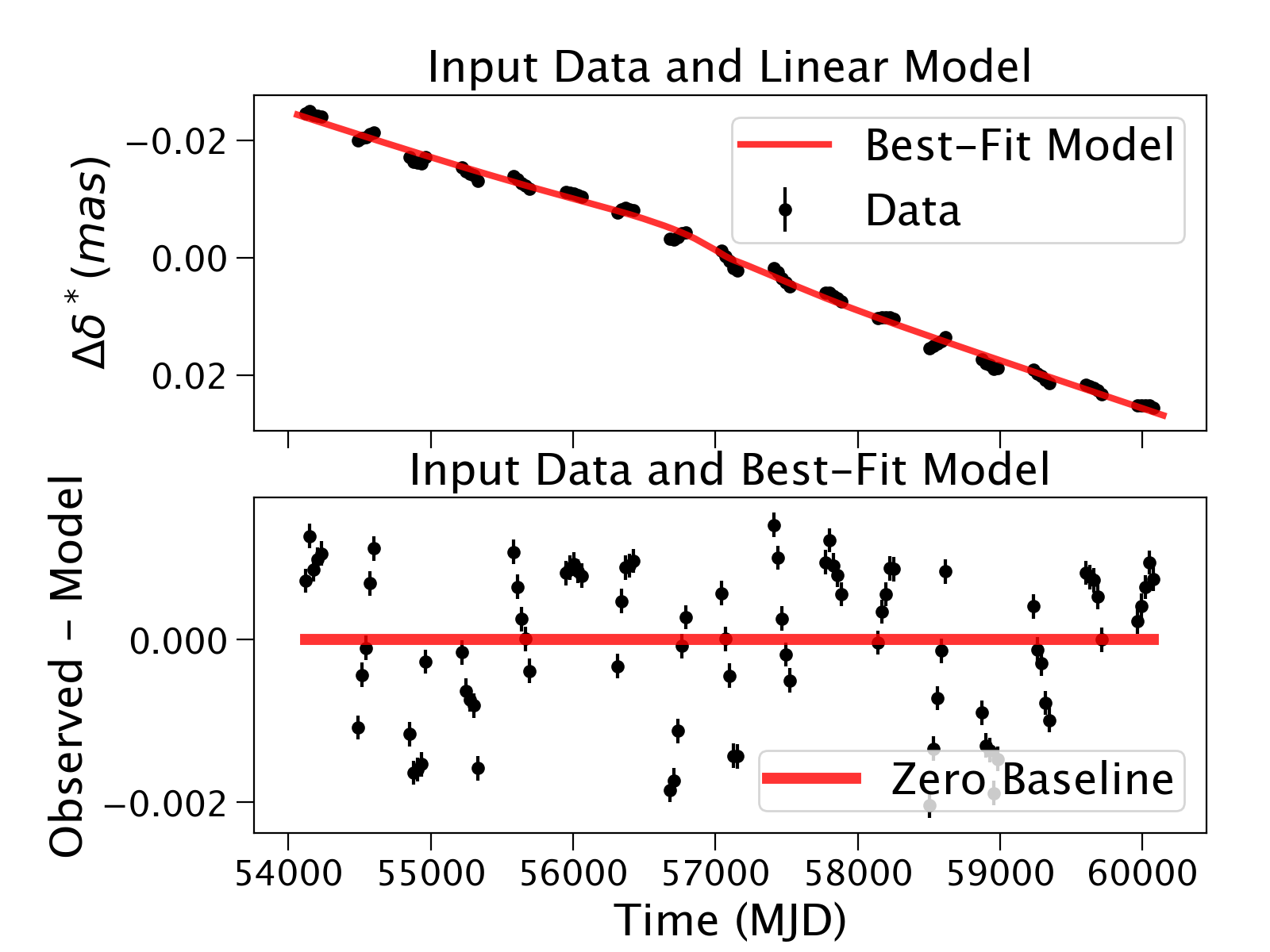}
    \includegraphics[width= .32\textwidth]{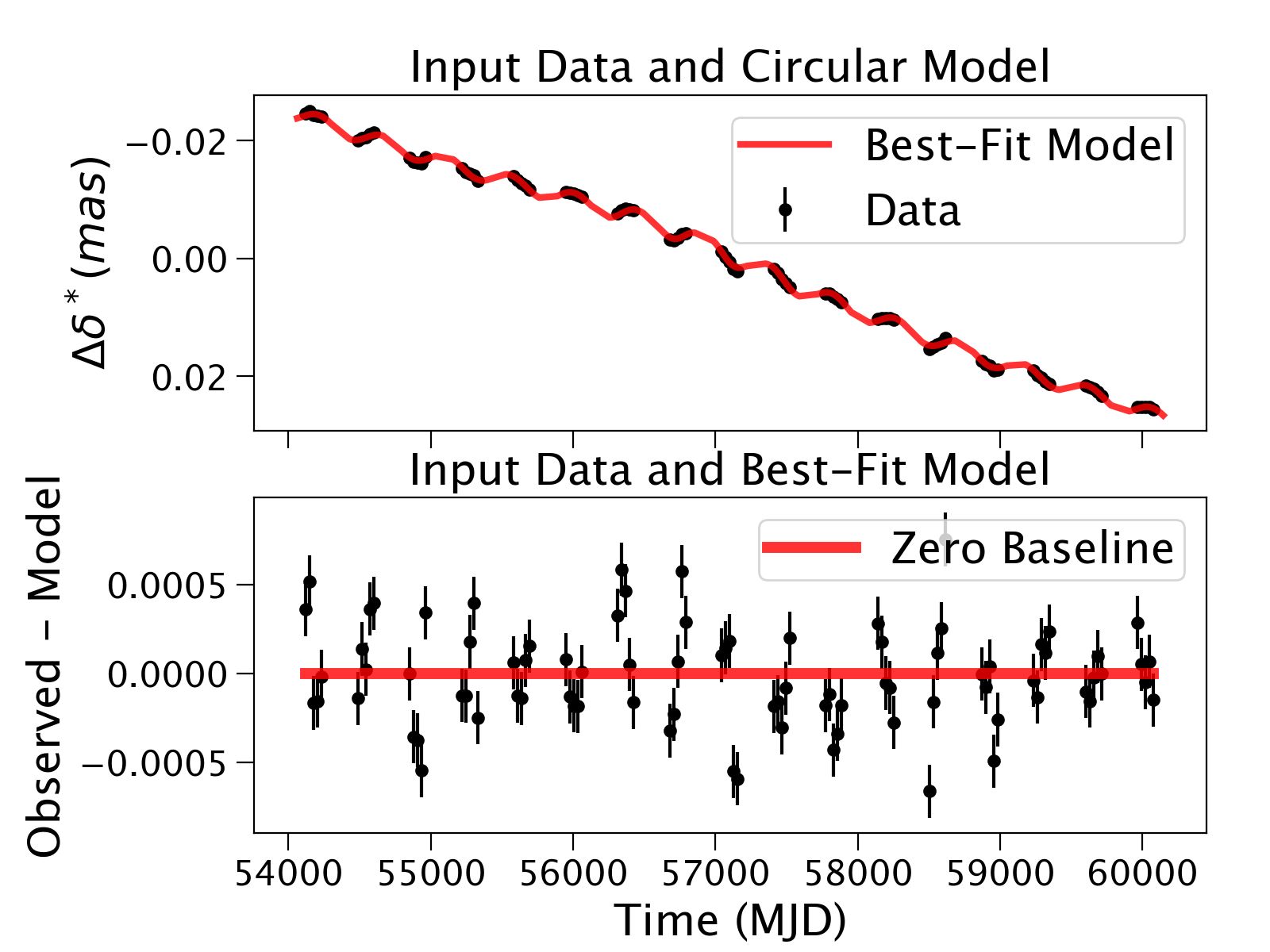}
    \includegraphics[width= .32\textwidth]{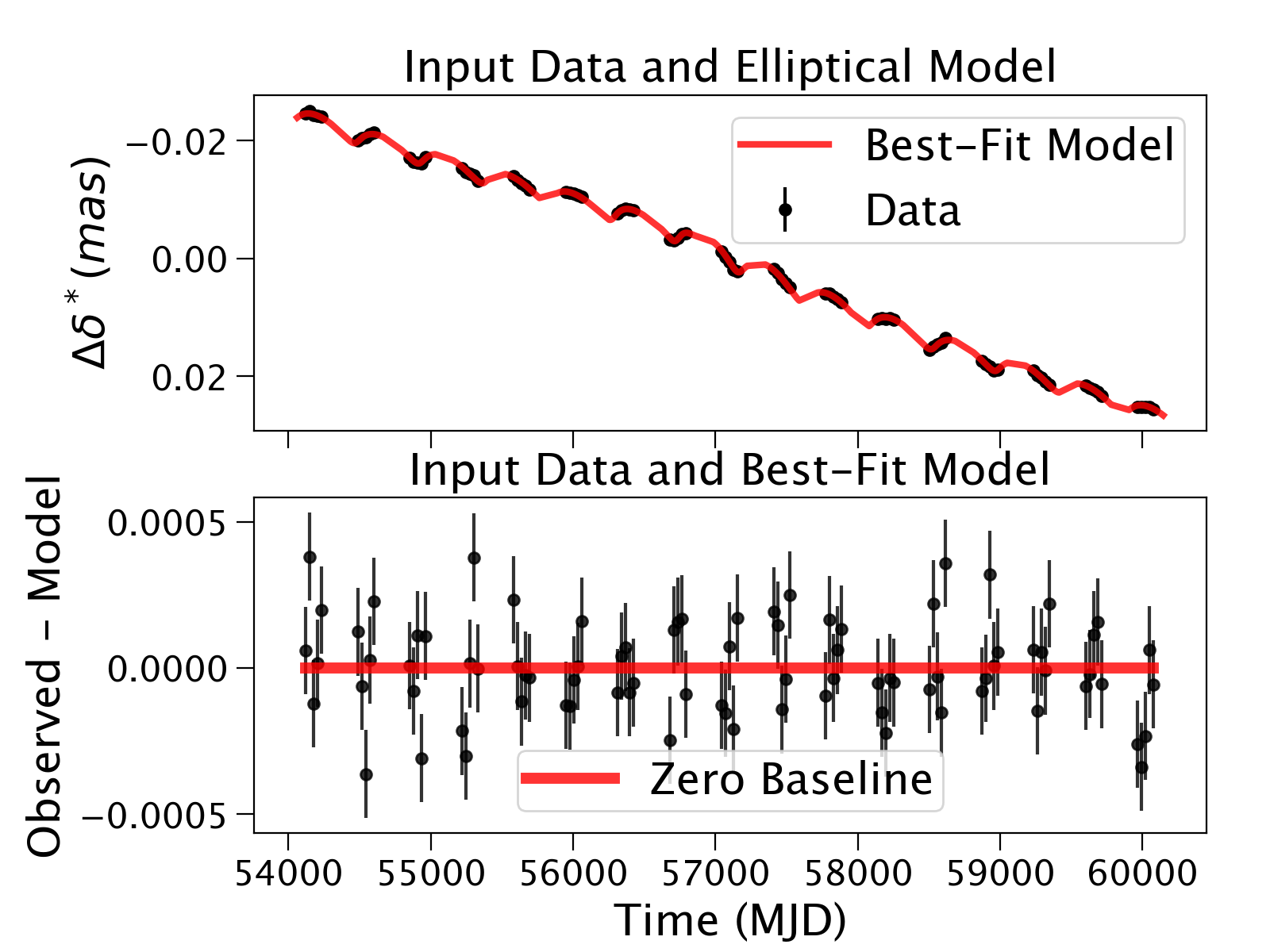}
    \caption{Fitting output for a mock astrometric dataset generated using  BSPL\_PhotAstrom\_noPar\_EllOrbs\_Param1 and the parameters displayed in Table~\ref{tab:fake_fit}. We present a fit for the Dec component in this figure. 
    (\emph{Left}) Best-fit with linear orbital motion. (\emph{Center}) Best-fit with circular orbital motion. (\emph{Right}) Best-fit with elliptical orbital motion.}
    \label{fig:orbital_comparison_as2}
\end{figure*}

\subsection{Results: Dependency of Fitting on Orbital Motion}
\label{sec:results_om}

\begin{deluxetable}{lc}
\tablecaption{Parameters used to generate the mock dataset using \texttt{BSPL\_PhotAstrom\_noPar\_EllOrbs\_Param1}.}
\label{tab:fake_fit}
\tablehead{
\colhead{\textbf{Parameter}} & \colhead{\textbf{Value}}
}
\startdata
\texttt{$m_L$} & $20\,M_\odot$ \\
\texttt{$\tcomnot$} & 57000 MJD\\
\texttt{$\uocom$} & 0.75 \\
\texttt{$dL$} & $1000 \, pc$ \\
\texttt{$dS$} & $10000 \, pc$ \\
\texttt{$\Xsvec$} & [0, 0] \\
\texttt{$\mulsvec$} & [0, 0] $\frac{mas}{yr}$ \\
\texttt{$\mussysvec$} & [8, 3] $\frac{mas}{yr}$ \\
\texttt{$\omega$} & $30 \degree$ \\
\texttt{$\Omega$} & $10 \degree$ \\
\texttt{$\inclination$} & $0 \degree$ \\
\texttt{$\eccentricity$} & 0.5 \\
\texttt{$\period$} & 450 days \\
\texttt{$\al$} & 2 mas \\
\texttt{$\ala$} & 2.5 mas \\
\texttt{$b_{sff}$} & 1 \\
\texttt{$mag_{S,pri}$} & 18 \\
\texttt{$mag_{S,sec}$} & 20 \\
\enddata
\end{deluxetable}

In this section, we demonstrate the need to account for the orbital motion of binary systems by fitting a mock dataset to various models with orbital motion in BAGLE. 

The mock dataset generated is intentionally designed to replicate a BSPL event with a complex lightcurve structure with apparent Keplerian motion. It is generated using BSPL\_PhotAstrom\_noPar\_EllOrbs\_Param1 and the parameters presented in Table~\ref{tab:fake_fit}. The mock dataset simulates photometric observations every day and astrometric observations every twenty-eight days for the bulge observing window. Data points were sampled from a model over a 6000 day window with a cadence of 1 day and 28 days for photometry and astrometry, respectively. Random noise was added assuming a photometric SNR=20 ($\sigma_{mag}$=0.05) and an astrometric error of 0.15 mas at mag=19.


In our fit, BSPL models with linear, circular, and elliptical orbital motion are utilized to demonstrate how incorporating orbital motion enhances the quality of the fit. We present the photometric fitting results in Figure~\ref{fig:orbital_comparison}, the astrometric RA fitting results in Figure~\ref{fig:orbital_comparison_as1} and the astrometric Dec fitting results in Figure~\ref{fig:orbital_comparison_as2}. 

Visually, the residuals improve when fitting a model with either a circular or elliptical orbital motion (as compared to the linear approximation), which is more closely aligned with the true nature of the mock dataset. 

Furthermore, the reduced chi-squared ($\bar{\chi}^2$) values are summarized in Table~\ref{tab:chi}, and capture the quality of the fits. In our reduced chi-squared test, we calculated the degrees of freedom by subtracting the number of fitting parameters from the total number of astrometric and photometric data points. The best-fit model with linear orbital motion has a $\bar{\chi}^2 = 3.2$. On the other hand, the best fits with circular and elliptical orbital motion have $\bar{\chi}^2 = 1.3$ and $\bar{\chi}^2 = 1.0$; these values indicate that the models with circular and elliptical orbital motion significantly improve our fitting results, and the residual difference between observed and fitted data is almost consistent with the error variance for the elliptical orbit model.

\begin{deluxetable}{lc}
\tablecaption{$\bar{\chi}^2$ values for the joint photometric and astrometric fit run on a mock dataset using \texttt{BSPL\_PhotAstrom\_noPar\_EllOrbs\_Param1} and parameters from Table~\ref{tab:fake_fit}. 
\label{tab:chi}}
\tablehead{
\colhead{\textbf{Orbital Motion}} & \colhead{\boldmath{$\bar{\chi}^2$}}
}
\startdata
Linear  & 3.2 \\
Circular & 1.3 \\
Elliptical & 1.0 \\
\enddata
\end{deluxetable}

From our reduced chi-squared test in Table \ref{tab:chi}, we conclude that incorporating Keplerian orbital motion into BAGLE is necessary to create best-fit models that fit well for complex lightcurves.


\section{Conclusion}
\label{sec:conclusion}

In this paper, we introduce binary models in BAGLE. These binary models account for binary sources, binary lenses or both (with and without orbital motion). Binary models with orbital motion in BAGLE can be divided into four categories: linear, accelerated, circular and elliptical. Models with circular and elliptical motion depend on eight crucial Keplerian elements ($\w$, $\bigomega$, $\inclination$, $\eccentricity$, $\period$, $t_p$, $\al$, and $\ala$), and are better-suited for microlensing events where $\period \ll \tE$. On the other hand, models with linear and accelerated motion use fewer free parameters, making them computationally inexpensive and well-suited approximations for microlensing events where $\period \gg \tE$.

From our fitting procedure using a mock dataset that replicates a binary-source, point-lens event, we conclude that the inclusion of orbital motion in binary microlensing events helps model complex photometric lightcurves. In these mock fitting, the accuracy of our binary fits based on $\bar{\chi^2}$ values improves with orbital motion. 

BAGLE's capabilities for handling point-source, point-lens and finite-source, point-lens events are presented in \citet{Lu:2025}, where BAGLE was compared with other microlensing packages like VBMicrolensing, pyLIMA, and MulensModel in detail. This paper includes a brief comparison between the different microlensing packages for point-source, binary-lens and binary-source, point-lens events. For point-source, binary-lens events and binary-source, point-lens events, the residual difference in magnification between VBMicrolensing and BAGLE ranged around $10^{-4}$. Between MulensModel and BAGLE, the difference ranged around $10^{-11}$ for point-source, binary-lens events and $10^{-15}$ for binary-source, point-lens events.

BAGLE's model-fitting runtimes are comparable to VBMicrolensing and MulensModel for pre-instantiated binary-source, point-lens events. A large-scale simulation of multiple events is fastest with BAGLE for binary-source, point-lens events. BAGLE's dependence on a polynomial root solver affects its runtime (full and pre-instantiated) for point-source, binary-lens calculations. Our future work involves exploring ways to improve the efficiency of the root solver through the use of JAX. We also aim to develop a similar lightcurve comparison for binary-source, binary-lens models, and compare events with the inclusion of orbital motion. 




In conclusion, the wide array of models and parameterizations available in BAGLE make it suitable for a joint photometric and astrometric fitting of binary events, including binary-source, binary-lens events. BAGLE's new binary models will be used to work with data from the Vera C. Rubin Observatory, the Nancy Grace Roman Telescope, and other surveys. These new models will be used to better characterize measured microlensing signals of black hole astrometric candidates. These new models, which accurately capture the orbital dynamics of binary systems, will enhance our search for dark lenses, such as black holes, exoplanets and other intriguing candidates. 

\begin{acknowledgments}
    We thank David Bennett, Valerio Bozza, Etienne Bachelet, and Radek Poleski for comments on package comparisons. We thank the Roman Galactic Exoplanet Project Infrastructure Team (RGES-PIT) working groups on microlens modeling and astrometry for useful discussions.
    The authors acknowledge support from the National Science Foundation under grant No.~2108185, the Heising-Simons Foundation under grant No.~2022-3542, and the Association of Universities for Research in Astronomy Space Telescope Science Institute under grant No.~HST-GO-17081.004-A and HST-GO-16658.001-A.   
\end{acknowledgments}

\software{Numpy \citep{oliphant2006guide}, Matplotlib \citep{Hunter:2007}, Astropy \citep{astropy:2013,astropy:2018,astropy:2022}, SciPy \citep{2020SciPy-NMeth}}



\pagebreak 

\appendix
In Appendix~\ref{sec:rectangular}, we describe the method used to find the rectangular coordinates. In Appendix~\ref{sec:Thiele-Innes} we describe the Thiele-Innes constants. In Appendices~\ref{sec:u0} and \ref{sec:t0}, we present the coordinate transformation in binary microlensing for $u_0$ and $t_0$. For example, BAGLE support binary lens parameterizations where the input parameters are in a reference frame with either the primary (or more massive) lens or the center of mass at origin. The different frames change the closest approach distance and time and the conversion between the two frames is non-trivial. These transformations can be used for both binary lens and binary source and can transform to any point along the binary axis. In Appendix ~\ref{sec:u0}, we transform $u_0$ and in Appendix ~\ref{sec:t0}, we transform $t_0$. 


\section{Finding the rectangular coordinates}
\label{sec:rectangular}
We begin by calculating the mean, eccentric, and true anomalies using the Keplerian orbital parameters. This method is adopted from \citet{Koren_2016}. The mean anomaly as a function of time $\M$ is:

\begin{eqnarray}
    \label{mean_anomaly}
    \M = \frac{2 \pi}{\period} (t-t_p)
\end{eqnarray}

For circular orbits, the mean anomaly is the same as the true anomaly. 

The eccentric anomaly ($\E$) can be found using the mean anomaly and the eccentricity of the orbit by solving:
 
\begin{eqnarray}
    \label{ecc_anomaly}
    \E - \eccentricity \sin \E = \M
\end{eqnarray}

The true anomaly $\etanom$ is found using $\M$ and $\E$. 


\begin{equation}
    \label{true_anomaly}
    \etanom = 2 \arctan \left( \sqrt{\frac{1+\eccentricity}{1-\eccentricity}} \tan{\frac{\E}{2}} \right)
\end{equation}

The mean, eccentric, and true anomalies help us define the elliptical rectangular coordinates of a binary system's orbit: 

\begin{equation}
    \label{ell_rec_coord1}
    \textit{X(t)} = \cos \E - \eccentricity 
\end{equation}

\begin{equation}
    \label{ell_rec_coord2}
    \textit{Y(t)} =  \sqrt{1 - \eccentricity^2} \sin \E
\end{equation}

\section{Finding Thiele-Innes Constants}
\label{sec:Thiele-Innes}

We can find the Thiele-Innes Constants using Keplerian elements. The Thiele-Innes Constants for the primary object are:

\begin{eqnarray}
    \Apri &=& \al \left(\cos \w \cos \bigomega - \sin \w \sin \bigomega \right) \nonumber \\
    \Bpri &=& \al \left(\cos \w \sin \bigomega + \sin \w \cos \bigomega \right) \nonumber  \\
    \Cpri &=& \al \left(\sin \w \sin \inclination\right) \nonumber \\
    \Fpri &=& \al \left(-\sin \w \cos \bigomega - \cos \w \sin \bigomega \cos \inclination \right) \nonumber \\
    \Gpri &=& \al \left(-\sin \w \sin \bigomega + \cos \w \cos \bigomega \cos \inclination \right) \nonumber \\
    \Hpri &=& \al \left(\cos \w \sin \inclination \right) 
\end{eqnarray}

In how we define our orbital parameterization, the only parameters we vary between the primary and secondary celestial objects are the length of the semi-major axis ($\al$ and $\ala$) and the argument of periastron ($\wsec = \w + 180 \degree$). Therefore, the Thiele-Innes Constants for the secondary object are

\begin{eqnarray}
    \Asec &=& \ala \left(\cos \wsec \cos \bigomega - \sin\wsec  \sin \bigomega \right) \nonumber \\
    \Bsec &=& \ala \left(\cos\wsec  \sin \bigomega + \sin\wsec  \cos \bigomega \right) \nonumber \\
    \Csec &=& \ala \left(\sin\wsec  \sin \inclination\right) \nonumber \\
    \Fsec &=& \ala \left(-\sin\wsec  \cos \bigomega - \cos\wsec  \sin \bigomega \cos \inclination \right) \nonumber  \\
    \Gsec &=& \ala \left(-\sin \wsec  \sin \bigomega + \cos\wsec  \cos \bigomega \cos \inclination \right) \nonumber  \\
    \Hsec &=& \ala \left(\cos \wsec \sin \inclination \right) 
\end{eqnarray}

\section{$\lowercase{u}_0$ transformation}
\label{sec:u0}
\begin{figure}[h]
    \centering
    \includegraphics[scale=0.3]{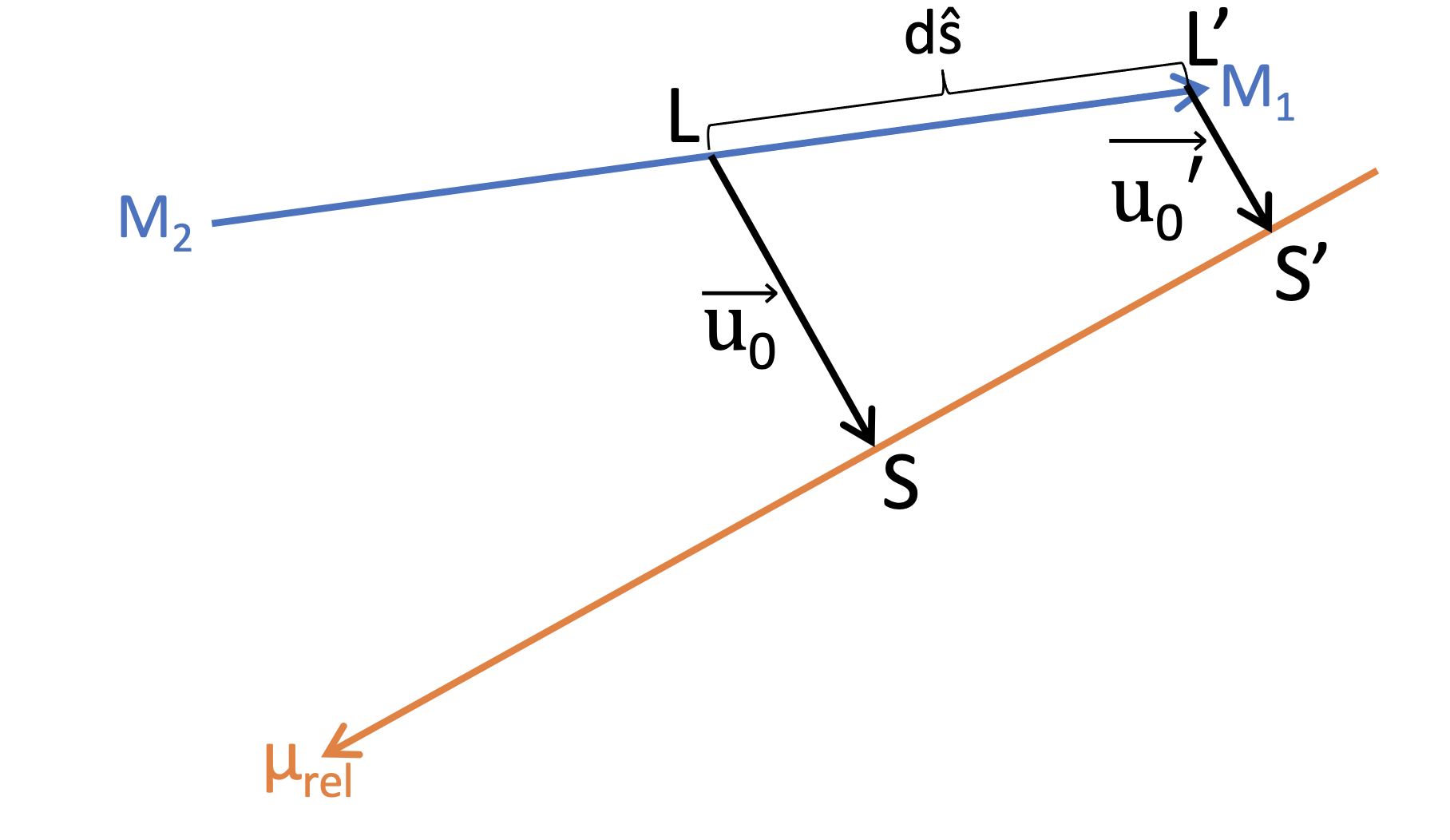}
    \caption{PSBL geometry where blue is the binary axis and orange is the source moving by with relative proper motion $\mu_{rel}$. We are transforming from $L$ to $L'$ where $L'$ is further along in the direction of $\hat{s}$ by $d$. The source's point of closest approach to $L$ is $S$ at a distance $u_0$ at time $t_0$. The source's point of closest approach to $L'$ is $S'$ at a distance $u_0'$ at time $t_0'$.}
    \label{fig:psbl coord transform}
\end{figure}

Transforming $u_0$ can be treated as a coordinate transformation of the origin from one point along the binary axis to another point (see Figure  ~\ref{fig:psbl coord transform}). We can transform the origin of the coordinate system from the geometric midpoint to the center of mass, to the primary, or to any other point. We preform the transformation in the heliocentric frame, so parallax does not need to be explicitly taken into account. Note that this coordinate transformation does not take orbital motion of the lens or source into account. It assumes that the angle of the binary axis and North and the direction of relative proper motion of the lens and source stay constant. $L$ is the initial position of the origin on the binary axis and $L'$ is the final position of the origin on the binary axis. $S$ is the closest the source gets to $L$ which occurs at time $t_0$ and $S'$ is the closest the source gets to $L'$ which occurs at time $t_0'$. The source is moving with a velocity $\vect{\mu_{rel}}$ (in the frame of the lens). We define a coordinate system $R$ with $L$ at the origin:
\begin{eqnarray}
    L &=& [0, 0] \\
    S &=& [u_{0, E}, u_{0, N}] \\
    L' &=& [d\hat{s}_E, d\hat{s}_N] \\
    S' &=& S + \frac{\vect{\mu_{rel}}}{\theta_E}(t_0' - t_0) = [u_{0, E} + \frac{\mu_{rel, E}}{\theta_E}(t_0' - t_0), u_{0, N} + \frac{\mu_{rel, N}}{\theta_E}(t_0' - t_0)]
\end{eqnarray}
where $\vect{d}$ is the vector in units of $\theta_E$ along the binary axis that we transform. The magnitude of the vector is $d$ and the direction is $\hat{s}$ where $\hat{s}_E$ and $\hat{s}_N$ are the East and North components. We normalize the proper motions by $\theta_E$ since $\vect{u_0}$ is in units of $\theta_E$.

We then define another coordinate system $R'$ with $L'$ at the origin:
\begin{eqnarray}
   L &=& [-d\hat{s}_E, -d\hat{s}_N] \\
    S &=& S' + \frac{\vect{\mu_{rel}}}{\theta_E}(t_0 - t_0') = [u_{0, E} + \frac{\mu_{rel, E}}{\theta_E}(t_0 - t_0'), u_{0, N} + \frac{\mu_{rel, N}}{\theta_E}(t_0 - t_0')] \\
    L' &=&  [0, 0] \\
    S' &=& [u_{0, E}', u_{0, N}'] 
\end{eqnarray}
So when we transform from $R \rightarrow R'$ we subtract $d\hat{s}$ since we shift the origin from $[0, 0]$ to $[d\hat{s}_E, d\hat{s}_N]$:
\begin{eqnarray}
    L^{R'} &=& L^R - d\hat{s} \\
    S^{R'} &=& S^R - d\hat{s} \\
    L'^{R'} &=& L'^R - d\hat{s} \\
    S'^{R'} &=& S'^R - d\hat{s}
\end{eqnarray}
We're interested in $\vect{u}_0'$ in terms of $\vect{u}_0$.
\begin{eqnarray}
    \vect{u}_0' &=& S'^{R'} - L'^{R'} = (S'^R - d\hat{s}) - (L'^R - d\hat{s}) = S'^R - L'^R \\
    \label{eq: u0' vectors}
    \vect{u}_0' &=& \vect{u}_0 + \frac{\vect{\mu}_{rel}}{\theta_E}(t_0' - t_0) - d\hat{s}
\end{eqnarray}
Let's break this into components where:
\begin{eqnarray}
    \vect{\mu}_{rel} &=& [\mu_{rel}\cos\theta_{\mu}, \mu_{rel}\sin\theta_{\mu}] \\
    \hat{s} &=& [\cos\theta_{s}, \sin\theta_{s}]
\end{eqnarray}
Where $\theta_{\mu}$ is the angle from North to $\vect{\mu}_{rel}$ and $\theta_s$ is the angle from North to the binary axis, East of North. (Note that choosing North as our reference will not affect the final answer. Another reference could be chosen). The binary axis is a vector from $M_2$ to $M_1$ as indicated in Figure~\ref{fig:psbl coord transform}.
So in components:
\begin{eqnarray}
\label{eq: u0', x}
    u_{0, E}' &=& u_{0, E} + \mu_{rel}\cos\theta_{\mu}\Big(\frac{t_0' - t_0}{\theta_E}\Big) - d\cos\theta_s \\
\label{eq: u0', y}
    u_{0, N}' &=& u_{0, N} + \mu_{rel}\sin\theta_{\mu}\Big(\frac{t_0' - t_0}{\theta_E}\Big) - d\sin\theta_s
\end{eqnarray}

The angle between $\vect{\mu}_{rel}$ and the binary axis ($\vect{s}$) is
\begin{equation}
    \phi = \theta_s - \theta_{\mu}.
\end{equation}
We will also be concerned with the angle to $\vect{u}_0$ from North ($\theta_u$). By definition it is always 90$^{\circ}$ off from $\theta_{\mu}$, but it is sometimes +90$^{\circ}$ and sometimes -90$^{\circ}$. We can find this sign by taking the cross product of $\hat{\mu}_{rel} \times \hat{u}_0$ and dotting the result with $\hat{z}$. $\hat{z}$ is a positive unit vector into the page.
\begin{equation}
\label{eq: C}
    C \equiv (\hat{\mu}_{rel} \times \hat{u}_0) \cdot \hat{z}
\end{equation}
where $C$ is -1 or 1. Since $\hat{z}$ is positive into the page, we subtract the result of the cross product
\begin{equation}
\label{eq: theta_u to theta_mu}
    \theta_u = \theta_{\mu} - 90^{\circ}C
\end{equation}
We can now take Eqs~\ref{eq: u0', x} and \ref{eq: u0', y} and define the components as their magnitudes times cosine/sine of angles:
\begin{eqnarray}
\label{eq: u0'cos}
    u_{0}'\cos\theta_{u'} &=& u_{0}\cos\theta_u + \mu_{rel}\cos\theta_{\mu}\Big(\frac{t_0' - t_0}{\theta_E}\Big) - d\cos\theta_s \\
\label{eq: u0'sin}
    u_{0}'\sin\theta_{u'} &=& u_{0}\sin\theta_u + \mu_{rel}\sin\theta_{\mu}\Big(\frac{t_0' - t_0}{\theta_E}\Big) - d\sin\theta_s
\end{eqnarray}
Where $\theta_{u'}$ is the angle from North to $\mathbf{u_0'}$. $\mathbf{u_0}$ will always be parallel to $\mathbf{u_0'}$, but they may be opposite directions. So
\begin{equation}
    \theta_{u'} = \theta_u + 180^{\circ}F
\end{equation}
where $F = 0$ if $u_0'$ and $u_0$ are parallel and $F = 1$ if they are antiparallel. Hence
\begin{eqnarray}
    \sin\theta_{u'} &=& G\sin\theta_u \\
    \cos\theta_{u'} &=& G\cos\theta_u
\end{eqnarray}
where $G = -1$ if $F = 1$ and $G = 1$ if $F = 0$.
Plugging into Eqs \ref{eq: u0'cos} and \ref{eq: u0'sin}:
\begin{eqnarray}
    u_{0}'G - u_0 &=& \frac{1}{\cos\theta_{u}}\Big(\mu_{rel}\cos\theta_{\mu}\Big(\frac{t_0' - t_0}{\theta_E}\Big) - d\cos\theta_s\Big) \\
    u_{0}'G - u_0&=&\frac{1}{\sin\theta_{u}}\Big(\mu_{rel}\sin\theta_{\mu}\Big(\frac{t_0' - t_0}{\theta_E}\Big) - d\sin\theta_s\Big)
\end{eqnarray}
We can also simplify 
\begin{eqnarray}
    \cos\theta_u &=& \cos(\theta_\mu - 90^{\circ}C) = C\sin\theta_{\mu} \\
    \sin\theta_u &=& \sin(\theta_\mu - 90^{\circ}C) = -C\cos\theta_{\mu}
\end{eqnarray}
since $C$ is either -1 or 1. Plugging that in:
\begin{eqnarray}
\label{eq: u0' - u0 from x}
    u_{0}'G - u_0 &=& \frac{1}{C}\Big(\mu_{rel}\frac{1}{\tan\theta_{\mu}}\Big(\frac{t_0' - t_0}{\theta_E}\Big) - d\frac{\cos\theta_s}{\sin\theta_{\mu}}\Big) \\
    \label{eq: u0' - u0 from y}
    u_{0}'G - u_0&=&-\frac{1}{C}\Big(\mu_{rel}\tan\theta_{\mu}\Big(\frac{t_0' - t_0}{\theta_E}\Big) - d\frac{\sin\theta_s}{\cos\theta_{\mu}}\Big)
\end{eqnarray}
We can set these equal to simplify. Some useful identities we'll use are:
\begin{eqnarray}
\label{eq: sin(theta_s) convention}
    \sin\theta_s &=& \sin(\theta_{\mu} + \phi) = \cos\phi\sin\theta_{\mu} + \cos\theta_{\mu}\sin\phi \\
\label{eq: cos(theta_s) convention}
    \cos\theta_s &=& \cos(\theta_{\mu} + \phi) = \cos\phi\cos\theta_{\mu} - \sin\phi\sin\theta_{\mu} \\
    \rightarrow \frac{\sin\theta_s}{\cos\theta_{\mu}} &=& \cos\phi\tan\theta_{\mu} + \sin\phi \\
    \frac{\cos\theta_s}{\sin\theta_{\mu}} &=& \frac{\cos\phi}{\tan\theta_{\mu}} - \sin\phi \\
    \rightarrow \frac{\sin\theta_s}{\cos\theta_{\mu}} + \frac{\cos\theta_s}{\sin\theta_{\mu}} &=& \cos\phi\Big(\frac{1}{\tan\theta_{\mu}} + \tan\theta_{\mu}\Big),
\end{eqnarray}
Setting Eq \ref{eq: u0' - u0 from x} equal to Eq \ref{eq: u0' - u0 from y} and simplifying:
\begin{eqnarray}
    \mu_{rel}\Big(\frac{t_0' - t_0}{\theta_E}\Big)\Big(\frac{1}{\tan\theta_{\mu}} + \tan\theta_{\mu}\Big) &=& d\Big( \frac{\sin\theta_s}{\cos\theta_{\mu}} + \frac{\cos\theta_s}{\sin\theta_{\mu}}\Big) \\
    \label{eq: mu_rel, ts, phi relation}
    \mu_{rel}\Big(\frac{t_0' - t_0}{\theta_E}\Big) &=& d \cos\phi
\end{eqnarray}
We can now use this relation in Eqs \ref{eq: u0', x} and \ref{eq: u0', y}. Starting with the E-component:
\begin{eqnarray}
\label{eq: u0x'}
    u_{0, E}' &=& u_{0, E} + d\cos\theta_{\mu}\cos\phi - d\cos\theta_s \\
    u_{0, E}' &=& u_{0, E} + d(\cos\theta_{\mu}\cos\phi - (\cos\phi\cos\theta_{\mu} - \sin\theta_{\mu}\sin\phi))\\
\label{eq: u0x' simplified}
    u_{0, E}' &=& u_{0, E} + d\sin\theta_{\mu}\sin\phi
\end{eqnarray}
Similarly for the N-component:
\begin{eqnarray}
\label{eq: u0y'}
    u_{0, N}' &=& u_{0, N} + d\sin\theta_{\mu}\cos\phi - d\sin\theta_s \\
    u_{0, N}' &=& u_{0, N} + d(\sin\theta_{\mu}\cos\phi - (\cos\phi\sin\theta_{\mu} + \cos\theta_{\mu}\sin\phi)) \\
    \label{eq: u0y' simplified}
    u_{0, N}' &=& u_{0, N}  - d\cos\theta_{\mu}\sin\phi.
\end{eqnarray}
We can use Eq. \ref{eq: theta_u to theta_mu}:
\begin{eqnarray}
\label{sin theta_mu conversion}
    \sin\theta_{\mu} &=& \sin(\theta_u + 90^{\circ}C) = C\cos\theta_u = C\hat{u}_{0,E} \\
\label{cos theta_mu conversion}
    \cos\theta_{\mu} &=& \cos(\theta_u + 90^{\circ}C) = -C\sin\theta_u = -C\hat{u}_{0,N}
\end{eqnarray}
Hence:
\begin{eqnarray}
    u_{0, E}' &=& u_{0, E} +C\hat{u}_{0,E}d\sin\phi \\
    u_{0, N}' &=& u_{0, N} +C\hat{u}_{0,N}d\sin\phi
\end{eqnarray}
Putting those together:
\begin{equation}
\label{eq: u0 transform general}
    \boxed{\vect{u_0}' = \vect{u_0} + C\hat{u}_0d\sin\phi}.
\end{equation}

We may also want to go the opposite direction. To do so we can define an equivalent of Eq. \ref{eq: C} for $\mathbf{u_0'}$:
\begin{equation}
    C' \equiv (\hat{\mu}_{rel} \times \hat{u}_0') \cdot \hat{z}
\end{equation}
So Eqs. \ref{sin theta_mu conversion} and \ref{cos theta_mu conversion} become
\begin{eqnarray}
    \sin\theta_{\mu} &=& \sin(\theta_{u'} + 90^{\circ}C') = C'\cos\theta_{u'} = C'\hat{u}'_{0,E} \\
    \cos\theta_{\mu} &=& \cos(\theta_{u'} + 90^{\circ}C') = -C'\sin\theta_{u'} = -C'\hat{u}'_{0,N}
\end{eqnarray}
Hence
\begin{eqnarray}
    u_{0, E}' &=& u_{0, E} +C'\hat{u}'_{0,E}d\sin\phi \\
    u_{0, N}' &=& u_{0, N} +C'\hat{u}'_{0,N}d\sin\phi \\
    \vect{u}_0' &=& \vect{u}_0 + C'\hat{u}'_0d\sin\phi
\end{eqnarray}
So Eq. \ref{eq: u0 transform general} becomes
\begin{equation}
    \label{eq: u0' transform general}
    \boxed{\vect{u}_0 = \vect{u}_0' - C'\hat{u}'_0d\sin\phi}.
\end{equation}

\subsection{Standard Coordinate Transforms}
\label{sec: standard transforms}
\subsubsection{Between Geometric Midpoint and Primary}
The separation in mas between the two lenses is $\vect{a}$ pointing towards the primary. In units of $\theta_{E}$, it's $\vect{s} \equiv \frac{\vect{a}}{\theta_E}$. So the vector from the geometric midpoint to the primary is $\frac{\vect{s}}{2}$. Hence Eq. \ref{eq: u0 transform general} becomes
\begin{equation}
    \vect{u_{\textrm{prim},0}} = \vect{u_{\textrm{geom}, 0}} + C\hat{u}_{\textrm{geom}, 0}\frac{a}{2\theta_E}\sin\phi.
\end{equation}
When transforming from primary to geometric midpoint, Eq. \ref{eq: u0' transform general} becomes:
\begin{equation}
    \vect{u_{\textrm{geom},0}} = \vect{u_{\textrm{prim},0}} - C'\hat{u}_{\textrm{prim}, 0}\frac{a}{2\theta_E}\sin\phi.
\end{equation}

\subsubsection{Between Geometric Midpoint and Center of Mass}
Following the derivation in ~\cite{Casey_Thesis}, Section 6.4.1, the separation between the geometric midpoint and center of mass in units of Einstein radii becomes:
\begin{equation}
    \vect{d} = \vect{s}\frac{1 - q}{2(1 + q)} \equiv \vect{s}q'.
\end{equation}
Hence Eq. \ref{eq: u0 transform general} becomes
\begin{equation}
    \vect{u_{\textrm{com}, 0}} = \vect{u_{ \textrm{geom}, 0}} + C\hat{u}_{\textrm{geom}, 0}sq'\sin\phi.
\end{equation}
When transforming from primary to geometric midpoint, Eq. \ref{eq: u0' transform general} becomes:
\begin{equation}
    \vect{u_{\textrm{geom},0}} = \vect{u_{\textrm{com},0}} - C'\hat{u}_{\textrm{com}, 0}sq'\sin\phi.
\end{equation}
If the secondary is more massive, then the center of mass is closer to the secondary than the primary, so the two equations will swtich.

\section{$\lowercase{t}_0$ transformation}
\label{sec:t0}

\begin{figure}
    \centering
    \includegraphics[scale=0.5]{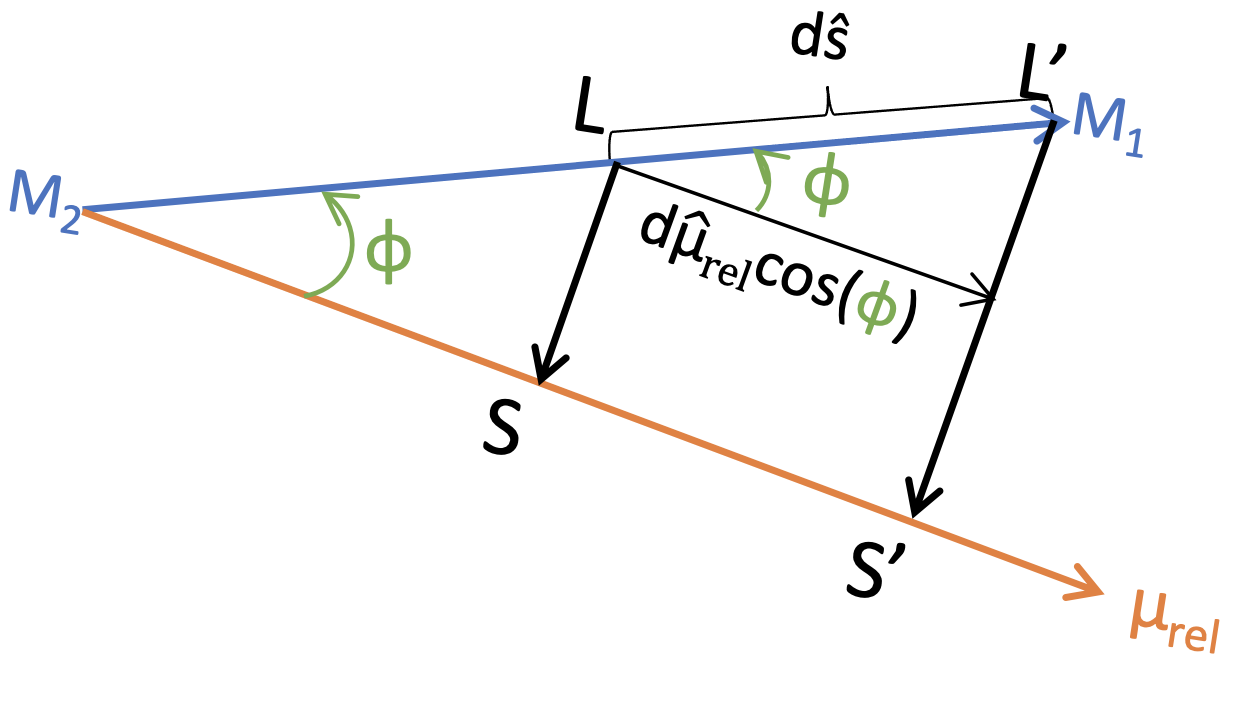}
    \caption{Similar to Figure~\ref{fig:psbl coord transform} but with the angle, $\phi$ between $\vect{\mu_{rel}}$ and the binary axis ($\vect{s}$) marked. $\phi < 90^{\circ}$ chosen for ease of visualization. The projection of $d\hat{s}$ onto $\vect{\mu_{rel}}$ is $\hat{\mu}_{rel}d\cos\phi$.}
    \label{fig:psbl geometry projection}
\end{figure}

Along with a change in the distance of closest approach, there is a change of when the closest approach occurs. In Figure~\ref{fig:psbl geometry projection}, the source is at $S$ at time $t_0$ and at $S'$ at time $t_0'$. Since the source is moving with relative proper motion $\boldsymbol{\mu_{rel}}$, we know:
\begin{equation}
    \label{eq: motion eq}
    S' - S = \frac{\vect{\mu_{rel}}}{\theta_E}(t_0' - t_0)
\end{equation}
The source moves across the Einstein radius ($\theta_E$) in time $t_E$, so:
\begin{equation}
    \mu_{rel} = \frac{\theta_E}{t_E}
\end{equation}
We can find $S' - S$ by projecting the binary axis onto $\vec{\mu}_{rel}$. The two are separated by angle $\phi$, so 
\begin{equation}
    S' - S = \hat{\mu}_{rel}d\cos\phi 
\end{equation}
Plugging this into Eq. \ref{eq: motion eq}, we find
\begin{eqnarray}
    \hat{\mu}_{rel}d\cos\phi  &=& \frac{\vect{\mu_{rel}}}{\theta_E}(t_0' - t_0) \\
    \hat{\mu}_{rel}d\cos\phi  &=& \frac{1}{t_E}(t_0' - t_0) \hat{\mu}_{rel}
\end{eqnarray}
\begin{equation}
\label{eq: t0 transform}
    \boxed{t_0' = t_0 + t_E d \cos\phi}
\end{equation}

\subsection{Standard Coordinate Transforms}
As described in Section \ref{sec: standard transforms}, for a geometric midpoint $\leftrightarrow$ primary center transformation Eq. \ref{eq: t0 transform} becomes
\begin{eqnarray}
    t_{prim, 0} = t_{geom, 0} + t_E\frac{a}{2\theta_E}\cos\phi
\end{eqnarray}
and for a geometric midpoint $\leftrightarrow$ center of mass transformation Eq. \ref{eq: t0 transform} becomes
\begin{eqnarray}
    t_{prim, 0}  = t_{geom, 0} + t_Esq' \cos\phi
\end{eqnarray}

\bibliographystyle{aasjournalv7}
\bibliography{main.bib}

\begin{thebibliography}{}
\expandafter\ifx\csname natexlab\endcsname\relax\def\natexlab#1{#1}\fi
\providecommand{\url}[1]{\href{#1}{#1}}
\providecommand{\dodoi}[1]{doi:~\href{http://doi.org/#1}{\nolinkurl{#1}}}
\providecommand{\doeprint}[1]{\href{http://ascl.net/#1}{\nolinkurl{http://ascl.net/#1}}}
\providecommand{\doarXiv}[1]{\href{https://arxiv.org/abs/#1}{\nolinkurl{https://arxiv.org/abs/#1}}}

\bibitem[{N.~S. {Abrams} {et~al.}(2025){Abrams}, {Lu}, {Lam}, {Medford}, {Hosek}, \& {Rose}}]{Natasha_2025}
{Abrams}, N.~S., {Lu}, J.~R., {Lam}, C.~Y., {et~al.} 2025, \bibinfo{title}{{Assessing the Impact of Binary Systems on Microlensing Using SPISEA and PopSyCLE Population Simulations},} \apj, 980, 103, \dodoi{10.3847/1538-4357/ada5f9}

\bibitem[{C. {Alcock} {et~al.}(2000){Alcock}, {Allsman}, {Alves}, {Axelrod}, {Baines}, {Becker}, {Bennett}, {Bourke}, {Brakel}, {Cook}, {Crook}, {Crouch}, {Dan}, {Drake}, {Fragile}, {Freeman}, {Gal-Yam}, {Geha}, {Gray}, {Griest}, {Gurtierrez}, {Heller}, {Howard}, {Johnson}, {Kaspi}, {Keane}, {Kovo}, {Leach}, {Leach}, {Leibowitz}, {Lehner}, {Lipkin}, {Maoz}, {Marshall}, {McDowell}, {McKeown}, {Mendelson}, {Messenger}, {Minniti}, {Nelson}, {Peterson}, {Popowski}, {Pozza}, {Purcell}, {Pratt}, {Quinn}, {Quinn}, {Rhie}, {Rodgers}, {Salmon}, {Shemmer}, {Stetson}, {Stubbs}, {Sutherland}, {Thomson}, {Tomaney}, {Vandehei}, {Walker}, {Ward}, \& {Wyper}}]{Alcock:2000-Binaries}
{Alcock}, C., {Allsman}, R.~A., {Alves}, D., {et~al.} 2000, \bibinfo{title}{{Binary Microlensing Events from the MACHO Project},} \apj, 541, 270, \dodoi{10.1086/309393}

\bibitem[{ {Astropy Collaboration} {et~al.}(2013){Astropy Collaboration}, {Robitaille}, {Tollerud}, {Greenfield}, {Droettboom}, {Bray}, {Aldcroft}, {Davis}, {Ginsburg}, {Price-Whelan}, {Kerzendorf}, {Conley}, {Crighton}, {Barbary}, {Muna}, {Ferguson}, {Grollier}, {Parikh}, {Nair}, {Unther}, {Deil}, {Woillez}, {Conseil}, {Kramer}, {Turner}, {Singer}, {Fox}, {Weaver}, {Zabalza}, {Edwards}, {Azalee Bostroem}, {Burke}, {Casey}, {Crawford}, {Dencheva}, {Ely}, {Jenness}, {Labrie}, {Lim}, {Pierfederici}, {Pontzen}, {Ptak}, {Refsdal}, {Servillat}, \& {Streicher}}]{astropy:2013}
{Astropy Collaboration}, {Robitaille}, T.~P., {Tollerud}, E.~J., {et~al.} 2013, \bibinfo{title}{{Astropy: A community Python package for astronomy},} \aap, 558, A33, \dodoi{10.1051/0004-6361/201322068}

\bibitem[{ {Astropy Collaboration} {et~al.}(2018){Astropy Collaboration}, {Price-Whelan}, {Sip{\H{o}}cz}, {G{\"u}nther}, {Lim}, {Crawford}, {Conseil}, {Shupe}, {Craig}, {Dencheva}, {Ginsburg}, {Vand erPlas}, {Bradley}, {P{\'e}rez-Su{\'a}rez}, {de Val-Borro}, {Aldcroft}, {Cruz}, {Robitaille}, {Tollerud}, {Ardelean}, {Babej}, {Bach}, {Bachetti}, {Bakanov}, {Bamford}, {Barentsen}, {Barmby}, {Baumbach}, {Berry}, {Biscani}, {Boquien}, {Bostroem}, {Bouma}, {Brammer}, {Bray}, {Breytenbach}, {Buddelmeijer}, {Burke}, {Calderone}, {Cano Rodr{\'\i}guez}, {Cara}, {Cardoso}, {Cheedella}, {Copin}, {Corrales}, {Crichton}, {D'Avella}, {Deil}, {Depagne}, {Dietrich}, {Donath}, {Droettboom}, {Earl}, {Erben}, {Fabbro}, {Ferreira}, {Finethy}, {Fox}, {Garrison}, {Gibbons}, {Goldstein}, {Gommers}, {Greco}, {Greenfield}, {Groener}, {Grollier}, {Hagen}, {Hirst}, {Homeier}, {Horton}, {Hosseinzadeh}, {Hu}, {Hunkeler}, {Ivezi{\'c}}, {Jain}, {Jenness}, {Kanarek}, {Kendrew}, {Kern}, {Kerzendorf}, {Khvalko}, {King}, {Kirkby}, {Kulkarni},
  {Kumar}, {Lee}, {Lenz}, {Littlefair}, {Ma}, {Macleod}, {Mastropietro}, {McCully}, {Montagnac}, {Morris}, {Mueller}, {Mumford}, {Muna}, {Murphy}, {Nelson}, {Nguyen}, {Ninan}, {N{\"o}the}, {Ogaz}, {Oh}, {Parejko}, {Parley}, {Pascual}, {Patil}, {Patil}, {Plunkett}, {Prochaska}, {Rastogi}, {Reddy Janga}, {Sabater}, {Sakurikar}, {Seifert}, {Sherbert}, {Sherwood-Taylor}, {Shih}, {Sick}, {Silbiger}, {Singanamalla}, {Singer}, {Sladen}, {Sooley}, {Sornarajah}, {Streicher}, {Teuben}, {Thomas}, {Tremblay}, {Turner}, {Terr{\'o}n}, {van Kerkwijk}, {de la Vega}, {Watkins}, {Weaver}, {Whitmore}, {Woillez}, {Zabalza}, \& {Astropy Contributors}}]{astropy:2018}
{Astropy Collaboration}, {Price-Whelan}, A.~M., {Sip{\H{o}}cz}, B.~M., {et~al.} 2018, \bibinfo{title}{{The Astropy Project: Building an Open-science Project and Status of the v2.0 Core Package},} \aj, 156, 123, \dodoi{10.3847/1538-3881/aabc4f}

\bibitem[{ {Astropy Collaboration} {et~al.}(2022){Astropy Collaboration}, {Price-Whelan}, {Lim}, {Earl}, {Starkman}, {Bradley}, {Shupe}, {Patil}, {Corrales}, {Brasseur}, {N{"o}the}, {Donath}, {Tollerud}, {Morris}, {Ginsburg}, {Vaher}, {Weaver}, {Tocknell}, {Jamieson}, {van Kerkwijk}, {Robitaille}, {Merry}, {Bachetti}, {G{"u}nther}, {Aldcroft}, {Alvarado-Montes}, {Archibald}, {B{'o}di}, {Bapat}, {Barentsen}, {Baz{'a}n}, {Biswas}, {Boquien}, {Burke}, {Cara}, {Cara}, {Conroy}, {Conseil}, {Craig}, {Cross}, {Cruz}, {D'Eugenio}, {Dencheva}, {Devillepoix}, {Dietrich}, {Eigenbrot}, {Erben}, {Ferreira}, {Foreman-Mackey}, {Fox}, {Freij}, {Garg}, {Geda}, {Glattly}, {Gondhalekar}, {Gordon}, {Grant}, {Greenfield}, {Groener}, {Guest}, {Gurovich}, {Handberg}, {Hart}, {Hatfield-Dodds}, {Homeier}, {Hosseinzadeh}, {Jenness}, {Jones}, {Joseph}, {Kalmbach}, {Karamehmetoglu}, {Ka{l}uszy{'n}ski}, {Kelley}, {Kern}, {Kerzendorf}, {Koch}, {Kulumani}, {Lee}, {Ly}, {Ma}, {MacBride}, {Maljaars}, {Muna}, {Murphy}, {Norman}, {O'Steen},
  {Oman}, {Pacifici}, {Pascual}, {Pascual-Granado}, {Patil}, {Perren}, {Pickering}, {Rastogi}, {Roulston}, {Ryan}, {Rykoff}, {Sabater}, {Sakurikar}, {Salgado}, {Sanghi}, {Saunders}, {Savchenko}, {Schwardt}, {Seifert-Eckert}, {Shih}, {Jain}, {Shukla}, {Sick}, {Simpson}, {Singanamalla}, {Singer}, {Singhal}, {Sinha}, {Sip{H{o}}cz}, {Spitler}, {Stansby}, {Streicher}, {{{S}}umak}, {Swinbank}, {Taranu}, {Tewary}, {Tremblay}, {Val-Borro}, {Van Kooten}, {Vasovi{'c}}, {Verma}, {de Miranda Cardoso}, {Williams}, {Wilson}, {Winkel}, {Wood-Vasey}, {Xue}, {Yoachim}, {Zhang}, {Zonca}, \& {Astropy Project Contributors}}]{astropy:2022}
{Astropy Collaboration}, {Price-Whelan}, A.~M., {Lim}, P.~L., {et~al.} 2022, \bibinfo{title}{{The Astropy Project: Sustaining and Growing a Community-oriented Open-source Project and the Latest Major Release (v5.0) of the Core Package},} \apj, 935, 167, \dodoi{10.3847/1538-4357/ac7c74}

\bibitem[{E. {Bachelet} {et~al.}(2017){Bachelet}, {Norbury}, {Bozza}, \& {Street}}]{Bachelet2017}
{Bachelet}, E., {Norbury}, M., {Bozza}, V., \& {Street}, R. 2017, \bibinfo{title}{{pyLIMA : an open source package for microlensing modeling. I. presentation of the software and analysis on single lens models},} arXiv e-prints, arXiv:1709.08704, \dodoi{10.48550/arXiv.1709.08704}

\bibitem[{D.~P. {Bennett}(2010){Bennett}}]{Bennett_2010}
{Bennett}, D.~P. 2010, \bibinfo{title}{{An Efficient Method for Modeling High-magnification Planetary Microlensing Events},} \apj, 716, 1408, \dodoi{10.1088/0004-637X/716/2/1408}

\bibitem[{V. {Bozza}(2010){Bozza}}]{Bozza2010}
{Bozza}, V. 2010, \bibinfo{title}{{Microlensing with an advanced contour integration algorithm: Green's theorem to third order, error control, optimal sampling and limb darkening},} \mnras, 408, 2188, \dodoi{10.1111/j.1365-2966.2010.17265.x}

\bibitem[{V. {Bozza}(2024){Bozza}}]{BozzaRT:2024}
{Bozza}, V. 2024, \bibinfo{title}{{RTModel: A platform for real-time modeling and massive analyses of microlensing events},} \aap, 688, A83, \dodoi{10.1051/0004-6361/202450450}

\bibitem[{V. {Bozza} {et~al.}(2018){Bozza}, {Bachelet}, {Bartoli{\'c}}, {Heintz}, {Hoag}, \& {Hundertmark}}]{Bozza2018}
{Bozza}, V., {Bachelet}, E., {Bartoli{\'c}}, F., {et~al.} 2018, \bibinfo{title}{{VBBINARYLENSING: a public package for microlensing light-curve computation},} \mnras, 479, 5157, \dodoi{10.1093/mnras/sty1791}

\bibitem[{V. {Bozza} {et~al.}(2021){Bozza}, {Khalouei}, \& {Bachelet}}]{Bozza2021}
{Bozza}, V., {Khalouei}, E., \& {Bachelet}, E. 2021, \bibinfo{title}{{A public code for astrometric microlensing with contour integration},} \mnras, 505, 126, \dodoi{10.1093/mnras/stab1376}

\bibitem[{V. {Bozza} {et~al.}(2024){Bozza}, {Saggese}, {Covone}, {Rota}, \& {Zhang}}]{Bozza2024}
{Bozza}, V., {Saggese}, V., {Covone}, G., {Rota}, P., \& {Zhang}, J. 2024, \bibinfo{title}{{VBMicroLensing: three algorithms for multiple lensing with contour integration},} arXiv e-prints, arXiv:2410.13660, \dodoi{10.48550/arXiv.2410.13660}

\bibitem[{J. Bradbury {et~al.}(2018)Bradbury, Frostig, Hawkins, Johnson, Leary, Maclaurin, Necula, Paszke, Vander{P}las, Wanderman-{M}ilne, \& Zhang}]{jax2018github}
Bradbury, J., Frostig, R., Hawkins, P., {et~al.} 2018, {JAX}: composable transformations of {P}ython+{N}um{P}y programs, 0.3.13 \url{http://github.com/jax-ml/jax}

\bibitem[{M. Dominik(2007)Dominik}]{Dominik_2007}
Dominik, M. 2007, \bibinfo{title}{Adaptive contouring - an efficient way to calculate microlensing light curves of extended sources,} Monthly Notices of the Royal Astronomical Society, 377, 1679–1688, \dodoi{10.1111/j.1365-2966.2007.11728.x}

\bibitem[{B.~S. {Gaudi}(2012){Gaudi}}]{Gaudi2012}
{Gaudi}, B.~S. 2012, \bibinfo{title}{{Microlensing Surveys for Exoplanets},} \araa, 50, 411, \dodoi{10.1146/annurev-astro-081811-125518}

\bibitem[{K. {Griest} \& N. {Safizadeh}(1998){Griest} \& {Safizadeh}}]{Griest1998}
{Griest}, K., \& {Safizadeh}, N. 1998, \bibinfo{title}{{The Use of High-Magnification Microlensing Events in Discovering Extrasolar Planets},} \apj, 500, 37, \dodoi{10.1086/305729}

\bibitem[{J.~D. Hunter(2007)Hunter}]{Hunter:2007}
Hunter, J.~D. 2007, \bibinfo{title}{Matplotlib: A 2D graphics environment,} Computing in Science \& Engineering, 9, 90, \dodoi{10.1109/MCSE.2007.55}

\bibitem[{M. {Jaroszynski}(2002){Jaroszynski}}]{Jaroszynski:2002}
{Jaroszynski}, M. 2002, \bibinfo{title}{{Binary Lenses in OGLE-II 1997-1999 Database. A Preliminary Study},} \actaa, 52, 39, \dodoi{10.48550/arXiv.astro-ph/0203476}

\bibitem[{S.~C. {Koren} {et~al.}(2016){Koren}, {Blake}, {Dahn}, \& {Harris}}]{Koren_2016}
{Koren}, S.~C., {Blake}, C.~H., {Dahn}, C.~C., \& {Harris}, H.~C. 2016, \bibinfo{title}{{The Low-mass Astrometric Binary LSR 1610-0040},} \aj, 151, 57, \dodoi{10.3847/0004-6256/151/3/57}

\bibitem[{C.~Y. {Lam}(2023){Lam}}]{Casey_Thesis}
{Lam}, C.~Y. 2023, \bibinfo{title}{{Understanding the Galactic Black Hole Population With Gravitational Microlensing},} PhD thesis, University of California, Berkeley, Department of Astronomy

\bibitem[{C.~Y. {Lam} \& J.~R. {Lu}(2023){Lam} \& {Lu}}]{Lam:2023-OB110462}
{Lam}, C.~Y., \& {Lu}, J.~R. 2023, \bibinfo{title}{{A re-analysis of the isolated black hole candidate OGLE-2011-BLG-0462/MOA-2011-BLG-191},} arXiv e-prints, arXiv:2308.03302, \dodoi{10.48550/arXiv.2308.03302}

\bibitem[{C.~Y. {Lam} {et~al.}(2022){Lam}, {Lu}, {Udalski}, {Bond}, {Bennett}, {Skowron}, {Mr{\'o}z}, {Poleski}, {Sumi}, {Szyma{\'n}ski}, {Koz{\l}owski}, {Pietrukowicz}, {Soszy{\'n}ski}, {Ulaczyk}, {Wyrzykowski}, {Miyazaki}, {Suzuki}, {Koshimoto}, {Rattenbury}, {Hosek}, {Abe}, {Barry}, {Bhattacharya}, {Fukui}, {Fujii}, {Hirao}, {Itow}, {Kirikawa}, {Kondo}, {Matsubara}, {Matsumoto}, {Muraki}, {Olmschenk}, {Ranc}, {Okamura}, {Satoh}, {Silva}, {Toda}, {Tristram}, {Vandorou}, {Yama}, {Abrams}, {Agarwal}, {Rose}, \& {Terry}}]{Lam:2022}
{Lam}, C.~Y., {Lu}, J.~R., {Udalski}, A., {et~al.} 2022, \bibinfo{title}{{An Isolated Mass-gap Black Hole or Neutron Star Detected with Astrometric Microlensing},} \apjl, 933, L23, \dodoi{10.3847/2041-8213/ac7442}

\bibitem[{J. {Lu} {et~al.}(submitted){Lu}, {Medford}, {Lam}, {Bhadra}, {Huston}, {Abrams}, {Broadberry}, {Chen}, {Terry}, {Arredondo}, \& {Scharf}}]{Lu:2025}
{Lu}, J., {Medford}, M., {Lam}, C., {et~al.} submitted, \bibinfo{title}{The BAGLE Python Package for Bayesian Analysis of Gravitational Lensing Events,} AAS Journals

\bibitem[{P. {McGill} {et~al.}(2023){McGill}, {Anderson}, {Casertano}, {Sahu}, {Bergeron}, {Blouin}, {Dufour}, {Smith}, {Evans}, {Belokurov}, {Smart}, {Bellini}, {Calamida}, {Dominik}, {Kains}, {Kl{\"u}ter}, {Nielsen}, \& {Wambsganss}}]{McGill2023}
{McGill}, P., {Anderson}, J., {Casertano}, S., {et~al.} 2023, \bibinfo{title}{{First semi-empirical test of the white dwarf mass-radius relationship using a single white dwarf via astrometric microlensing},} \mnras, 520, 259, \dodoi{10.1093/mnras/stac3532}

\bibitem[{P. {Mr{\'o}z} {et~al.}(2022){Mr{\'o}z}, {Udalski}, \& {Gould}}]{Mroz:2022}
{Mr{\'o}z}, P., {Udalski}, A., \& {Gould}, A. 2022, \bibinfo{title}{{Systematic Errors as a Source of Mass Discrepancy in Black Hole Microlensing Event OGLE-2011-BLG-0462},} \apjl, 937, L24, \dodoi{10.3847/2041-8213/ac90bb}

\bibitem[{P. {Mr{\'o}z} {et~al.}(2019){Mr{\'o}z}, {Udalski}, {Skowron}, {Szyma{\'n}ski}, {Soszy{\'n}ski}, {Wyrzykowski}, {Pietrukowicz}, {Koz{\l}owski}, {Poleski}, {Ulaczyk}, {Rybicki}, \& {Iwanek}}]{Mroz:2019}
{Mr{\'o}z}, P., {Udalski}, A., {Skowron}, J., {et~al.} 2019, \bibinfo{title}{{Microlensing Optical Depth and Event Rate toward the Galactic Bulge from 8 yr of OGLE-IV Observations},} \apjs, 244, 29, \dodoi{10.3847/1538-4365/ab426b}

\bibitem[{P. Mróz \& R. Poleski(2024)Mróz \& Poleski}]{Mroz_2024}
Mróz, P., \& Poleski, R. 2024, Exoplanet Occurrence Rates from Microlensing Surveys (Springer International Publishing), 1–23, \dodoi{10.1007/978-3-319-30648-3_208-1}

\bibitem[{T.~E. Oliphant {et~al.}(2006)Oliphant {et~al.}}]{oliphant2006guide}
Oliphant, T.~E., {et~al.} 2006, Guide to numpy, Vol.~1 (Trelgol Publishing USA)

\bibitem[{R. {Poleski} \& J.~C. {Yee}(2019){Poleski} \& {Yee}}]{Poleski2019}
{Poleski}, R., \& {Yee}, J.~C. 2019, \bibinfo{title}{{Modeling microlensing events with MulensModel},} Astronomy and Computing, 26, 35, \dodoi{10.1016/j.ascom.2018.11.001}

\bibitem[{K.~C. {Sahu} {et~al.}(2017){Sahu}, {Anderson}, {Casertano}, {Bond}, {Bergeron}, {Nelan}, {Pueyo}, {Brown}, {Bellini}, {Levay}, {Sokol}, {Dominik}, {Calamida}, {Kains}, \& {Livio}}]{Sahu2017}
{Sahu}, K.~C., {Anderson}, J., {Casertano}, S., {et~al.} 2017, \bibinfo{title}{{Relativistic deflection of background starlight measures the mass of a nearby white dwarf star},} Science, 356, 1046, \dodoi{10.1126/science.aal2879}

\bibitem[{K.~C. {Sahu} {et~al.}(2022){Sahu}, {Anderson}, {Casertano}, {Bond}, {Udalski}, {Dominik}, {Calamida}, {Bellini}, {Brown}, {Rejkuba}, {Bajaj}, {Kains}, {Ferguson}, {Fryer}, {Yock}, {Mr{\'o}z}, {Koz{\l}owski}, {Pietrukowicz}, {Poleski}, {Skowron}, {Soszy{\'n}ski}, {Szyma{\'n}ski}, {Ulaczyk}, {Wyrzykowski}, {Barry}, {Bennett}, {Bond}, {Hirao}, {Silva}, {Kondo}, {Koshimoto}, {Ranc}, {Rattenbury}, {Sumi}, {Suzuki}, {Tristram}, {Vandorou}, {Beaulieu}, {Marquette}, {Cole}, {Fouqu{\'e}}, {Hill}, {Dieters}, {Coutures}, {Dominis-Prester}, {Bennett}, {Bachelet}, {Menzies}, {Albrow}, {Pollard}, {Gould}, {Yee}, {Allen}, {Almeida}, {Christie}, {Drummond}, {Gal-Yam}, {Gorbikov}, {Jablonski}, {Lee}, {Maoz}, {Manulis}, {McCormick}, {Natusch}, {Pogge}, {Shvartzvald}, {J{\o}rgensen}, {Alsubai}, {Andersen}, {Bozza}, {Novati}, {Burgdorf}, {Hinse}, {Hundertmark}, {Husser}, {Kerins}, {Longa-Pe{\~n}a}, {Mancini}, {Penny}, {Rahvar}, {Ricci}, {Sajadian}, {Skottfelt}, {Snodgrass}, {Southworth}, {Tregloan-Reed}, {Wambsganss},
  {Wertz}, {Tsapras}, {Street}, {Bramich}, {Horne}, {Steele}, \& {RoboNet Collaboration}}]{Sahu:2022}
{Sahu}, K.~C., {Anderson}, J., {Casertano}, S., {et~al.} 2022, \bibinfo{title}{{An Isolated Stellar-mass Black Hole Detected through Astrometric Microlensing},} \apj, 933, 83, \dodoi{10.3847/1538-4357/ac739e}

\bibitem[{K.~C. {Sahu} {et~al.}(2025){Sahu}, {Anderson}, {Casertano}, {Bond}, {Dominik}, {Calamida}, {Bellini}, {Brown}, {Ferguson}, \& {Rejkuba}}]{Sahu2025}
{Sahu}, K.~C., {Anderson}, J., {Casertano}, S., {et~al.} 2025, \bibinfo{title}{{OGLE-2011-BLG-0462: An Isolated Stellar-mass Black Hole Confirmed Using New HST Astrometry and Updated Photometry},} \apj, 983, 104, \dodoi{10.3847/1538-4357/adbe6e}

\bibitem[{P. {Schneider} \& A. {Weiss}(1986){Schneider} \& {Weiss}}]{Schneider_1986}
{Schneider}, P., \& {Weiss}, A. 1986, \bibinfo{title}{{The two-point-mass lens - Detailed investigation of a special asymmetric gravitational lens},} \aap, 164, 237

\bibitem[{T.~N. {Thiele}(1883){Thiele}}]{Thiele1883}
{Thiele}, T.~N. 1883, \bibinfo{title}{{Neue Methode zur Berechung von Doppelsternbahnen},} Astronomische Nachrichten, 104, 245

\bibitem[{P. Virtanen {et~al.}(2020)Virtanen, Gommers, Oliphant, Haberland, Reddy, Cournapeau, Burovski, Peterson, Weckesser, Bright, {van der Walt}, Brett, Wilson, Millman, Mayorov, Nelson, Jones, Kern, Larson, Carey, Polat, Feng, Moore, {VanderPlas}, Laxalde, Perktold, Cimrman, Henriksen, Quintero, Harris, Archibald, Ribeiro, Pedregosa, {van Mulbregt}, \& {SciPy 1.0 Contributors}}]{2020SciPy-NMeth}
Virtanen, P., Gommers, R., Oliphant, T.~E., {et~al.} 2020, \bibinfo{title}{{{SciPy} 1.0: Fundamental Algorithms for Scientific Computing in Python},} Nature Methods, 17, 261, \dodoi{10.1038/s41592-019-0686-2}

\bibitem[{H.~J. {Witt}(1990){Witt}}]{Witt1990}
{Witt}, H.~J. 1990, \bibinfo{title}{{Investigation of high amplification events in light curves of gravitationally lensed quasars.},} \aap, 236, 311

\bibitem[{H.~J. {Witt} \& S. {Mao}(1995){Witt} \& {Mao}}]{WittShude1995}
{Witt}, H.~J., \& {Mao}, S. 1995, \bibinfo{title}{{On the Minimum Magnification Between Caustic Crossings for Microlensing by Binary and Multiple Stars},} \apjl, 447, L105, \dodoi{10.1086/309566}

\end{thebibliography}

\end{document}